\documentclass[]{aa}

\usepackage{graphicx}
\usepackage{txfonts}
\usepackage[colorlinks=true,citecolor=blue,linkcolor=blue]{hyperref}
\usepackage{natbib}
\bibpunct{(}{)}{;}{a}{}{,} 
\usepackage{bm}
\usepackage{upgreek}
\usepackage[dvipsnames]{xcolor}
\usepackage{amsmath}    
\usepackage{orcidlink}

\newcommand{\Ho}{$H_0$}

\newcommand{\lensname}{RXJ1131$-$1231}
\newcommand{\jwst}{{JWST}}

\newcommand{\caii}{Ca \textsc{ii}}
\newcommand{\cat}{Ca \textsc{ii} triplet}

\newcommand{\hunit}{km s$^{-1}$ Mpc$^{-1}$}
\newcommand{\kmps}{km s$^{-1}$}

\newcommand{\secref}[1]{Sect.~\ref{#1}}
\newcommand{\figref}[1]{Fig.~\ref{#1}}

\newcommand{\appref}[1]{App.~\ref{#1}}

\newcommand{\shajiblsft}{\citet{Shajib25a}}
\newcommand{\shajibraccoon}{\citep{Shajib25b}}
\newcommand{\shajibraccoont}{\citet{Shajib25b}}

\begin{document} 

    \title{TDCOSMO}
    
    \subtitle{XXIV. First spatially resolved kinematics of the lens galaxy obtained using JWST-NIRSpec to improve time-delay cosmography
    }
    
    \titlerunning{Spatially resolved kinematics of \lensname\ from \jwst}
    \authorrunning{A.~J.~Shajib et al.}
    
    \author{
    	Anowar~J.~Shajib\orcidlink{0000-0002-5558-888X},\inst{\ref{uchicago},\ref{kicp},\ref{cassa}}\fnmsep\thanks{NFHP Einstein Fellow}
            Tommaso~Treu\orcidlink{0000-0002-8460-0390},\inst{\ref{ucla}}
   		Sherry~H.~Suyu\orcidlink{0000-0001-5568-6052},\inst{\ref{tum},\ref{mpa}}
        David Law\orcidlink{0000-0002-9402-186X},\inst{\ref{stsci}}
            Ak{\i}n Y{\i}ld{\i}r{\i}m,\inst{\ref{mpa}} 
   		Michele~Cappellari\orcidlink{0000-0002-1283-8420},\inst{\ref{oxford}}
        Aymeric~Galan\orcidlink{0000-0003-2547-9815},\inst{\ref{tum},\ref{mpa}}
	    Shawn~Knabel\orcidlink{0000-0001-5110-6241},\inst{\ref{ucla}}
        Han~Wang\orcidlink{0000-0002-1293-5503},\inst{\ref{tum},\ref{mpa}}
            Simon~Birrer\orcidlink{0000-0003-3195-5507},\inst{\ref{stonybrook}}
    Fr\'ed\'eric~Courbin\orcidlink{0000-0003-0758-6510},\inst{\ref{iccub},\ref{icrea}}
    Christopher~D.~Fassnacht\orcidlink{0000-0002-4030-5461},\inst{\ref{ucd}}
            Joshua~A.~Frieman\orcidlink{0000-0003-4079-3263},\inst{\ref{uchicago},\ref{kicp},\ref{slac}}
            Alejandra Melo\orcidlink{0000-0002-6449-3970},\inst{\ref{mpa}, \ref{tum}}
            Takahiro~Morishita\orcidlink{0000-0002-8512-1404},\inst{\ref{ipac}}
            Pritom~Mozumdar\orcidlink{0000-0002-8593-7243},\inst{\ref{ucla}}
    	Dominique~Sluse\orcidlink{0000-0001-6116-2095},\inst{\ref{star}} and
        Massimo~Stiavelli\orcidlink{0000-0001-9935-6047}\inst{\ref{stsci}}
    }
    
    \institute{
    Department  of  Astronomy  \&  Astrophysics,  University  of Chicago, Chicago, IL 60637, USA; \email{ajshajib@uchicago.edu} \label{uchicago}
    \and
    Kavli Institute for Cosmological Physics, University of Chicago, Chicago, IL 60637, USA \label{kicp}
    \and
    Center for Astronomy, Space Science and Astrophysics, Independent University, Bangladesh, Dhaka 1229, Bangladesh \label{cassa}
    \and
    Department of Physics and Astronomy, University of California, Los Angeles, CA 90095, USA \label{ucla}
    \and
    Technical University of Munich, TUM School of Natural Sciences, Physics Department, James-Franck-Str.~1, Garching, 85748, Germany \label{tum}
    \and
    Max Planck Institute for Astrophysics, Karl-Schwarzschild-Str.~1, Garching, 85748, Germany \label{mpa}
    \and
    Space Telescope Science Institute, 3700 San Martin Dr., Baltimore, MD 21218, USA \label{stsci}
    \and
    Sub-Department of Astrophysics, Department of Physics, University of Oxford, Denys Wilkinson Building, Keble Road, Oxford, OX1 3RH, UK \label{oxford}
    \and
    Department of Physics and Astronomy, Stony Brook University, Stony Brook, NY 11794, USA \label{stonybrook}
    \and 
    Institut de Ci\`{e}ncies del Cosmos (ICCUB), 
    Universitat de Barcelona (IEEC-UB), 
    Mart\'{i} i Franqu\`{e}s 1, 08028 Barcelona, Spain \label{iccub}
    \and
    Instituci\'o Catalana de Recerca i Estudis Avan\c{c}ats (ICREA), 
    Passeig de Llu\'{i}s Companys 23, 08010 Barcelona, Spain \label{icrea}
       \and
    Department of Physics and Astronomy, UC Davis, 1 Shields Ave., Davis, CA 95616, USA \label{ucd}
    \and
    SLAC National Laboratory, 2575 Sand Hill Rd, Menlo Park, CA 94025 \label{slac}
        \and
    IPAC, California Institute of Technology, MC 314-6, 1200 E. California Boulevard, Pasadena, CA 91125, USA \label{ipac}
    \and
    STAR Institute, Li\`ege Universit\'e, Quartier Agora - All\'ee du six Ao\^ut, 19c B-4000 Li\`ege, Belgium \label{star}
    }
             
   \date{Received xxx, xxxx; accepted xxx, xxxx}

 
  \abstract
  {Spatially resolved stellar kinematics has become a key ingredient in time-delay cosmography to break the mass-sheet degeneracy in the mass profile and in turn provide a precise constraint on the Hubble constant and other cosmological parameters. In this paper, we present the first measurements of 2D resolved stellar kinematics for the lens galaxy in the quadruply lensed quasar system \lensname\ using integral field spectroscopy from JWST's Near-Infrared Spectrograph (NIRSpec), marking the first such measurement conducted with JWST. In extracting robust kinematic measurements from this first-of-its-kind dataset, we have made methodological improvements both in the data reduction and kinematic extraction. In our kinematic extraction procedure, we performed joint modeling of the lens galaxy, the quasar, and its host galaxy's contributions in the spectra to deblend the lens galaxy component and robustly constrain its stellar kinematics. Our improved methodological frameworks are released as software pipelines for future use: \textsc{squirrel}, for extracting stellar kinematics, and \textsc{RegalJumper}, for JWST-NIRSpec data reduction. We incorporated additional artifact cleaning beyond the standard JWST pipeline. We compared our measured stellar kinematics from the JWST NIRSpec with previously obtained ground-based measurements from the Keck Cosmic Web Imager integral field unit and find that the two datasets are statistically consistent at a $\sim$1.1$\sigma$ confidence level. Our measured kinematics will be used in a future study to improve the precision of the Hubble constant measurement.
}
  

  \maketitle
%
\section{Introduction}

The so-called Hubble tension -- the statistical difference between the Hubble constant, \Ho, measured by several late-Universe probes and that inferred from the early-Universe probes \citep{Abdalla22, DiValentino25} -- is arguably the most pressing issue in observational cosmology. If the tension is confirmed to be real and not due to unknown systematic uncertainties in multiple measurements, solving it would require new physics beyond the standard flat Lambda cold dark matter cosmology. For example, one class of solutions would require modifying the expansion history before recombination to adjust the sound horizon at recombination, either by introducing new relativistic particles or early dark energy \citep{Knox19}. Other solutions are also possible. However, it is not trivial to find a solution that solves the tension while satisfying all the other cosmological probes \citep{Vagnozzi23}, even with models featuring a late-time evolution in the dark energy favored by recent measurements \citep{Lynch24}.

Time-delay cosmography \citep{Treu16, Treu22, Treu23, Birrer24} measures the Hubble constant and other cosmological parameters in a way that is complementary and fully independent of traditional methods. It is a "local" measurement in the sense that it measures distances out to the lensed sources (typically $z\sim1$--3), thus comparable to the local distance ladder, but it reaches the Hubble flow directly in one step. Furthermore, by measuring the absolute time-delay distance, which is a ratio of angular diameter distances, its degeneracies are complementary to those of other probes \citep{Linder11}. As a cosmological probe, it is most similar to baryonic acoustic oscillations, but it does not require knowledge of the sound horizon at recombination for absolute calibration.

As the culmination of two decades of effort, in 2020 the Time-Delay COSMOgraphy (TDCOSMO) collaboration and its forerunners \citep[COSMOGRAIL, H0LiCOW, SHARP, STRIDES;][]{Eigenbrod05, Lagattuta12, Suyu17, Treu18} published a measurement of the Hubble constant with 2\% precision based on seven lenses \citep[][hereafter, TDCOSMO-I]{Millon20a}. The measurement agreed very well with the local distance ladder measurement by the Supernova \Ho\ for the Equation of State (SH0ES) team \citep{Riess22}, strengthening the tension with the early-Universe probe.

The TDCOSMO-I measurement achieved its precision by breaking the mass-sheet degeneracy \citep[MSD;][]{Falco85} implicitly via the assumption of simply parameterized mass density profiles for the deflector galaxies. These models -- power-law mass density profile, or stars plus Navarro--Frenk--White dark matter density profiles \citep{Navarro97} -- had been shown to describe the mass profiles of elliptical galaxies by studies based on stellar kinematics \citep{Cappellari13} and X-ray measurements \citep{Humphrey10}.

Given the importance of the tension, the TDCOSMO collaboration decided to repeat the measurements under virtually no assumptions about the mass density profile. By choosing a parameterization of the mass density profile that is maximally degenerate with \Ho\ via the MSD, \citet[][TDCOSMO-IV]{Birrer20} showed that the uncertainties increase from 2\% to 9\% for the same systems, where the information content of the unresolved stellar kinematics available at that time was the limiting factor. \citet{TDCOSMO25} provide the most recent measurement from time-delay cosmography, increasing the sample size of time-delay lenses to eight and improving, among other aspects, all the stellar kinematic measurements from new datasets, leading to a 4--5\% measurement of the Hubble constant ($H_0 = 71.6^{+3.9}_{-3.3}$ \hunit\ for the TDCOSMO lenses in combination with a prior on $\Omega_{\rm m}$ from Pantheon+ supernovae and $74.8_{-3.4}^{+3.5}$ \hunit\ from TDCOSMO+SLACS+SL2S in combination with a prior on $\Omega_{\rm m}$ from the DESI baryon acoustic oscillation).\footnote{SLACS: Sloan Lens ACS \citep{Bolton06}; SL2S: Strong Lensing Legacy Survey \citep{Gavazzi12}; DESI: Dark Energy Spectroscopic Instrument \citep{DESI25}.} Stellar kinematics, in particular spatially resolved information, is a crucial element in further improving the precision of time-delay cosmography under these weaker assumptions, as it breaks the MSD and the mass-anisotropy degeneracy \citep[MAD;][]{Binney82, Treu02a,Shajib18, Birrer20, Birrer21, Yildirim20, Yildirim23}.

\citet{Shajib23} have presented a spatially resolved velocity dispersion measurement of \lensname, the first of its kind. This measurement is based on integral field spectroscopy (IFS) from the Keck Observatory's Keck Cosmic Web Imager (KCWI) instrument. In combination with lens models and time delays previously published by our collaboration \citep{Tewes13, Suyu13, Suyu14}, \citet{Shajib23} measured $H_0 = 77.1_{-7.1}^{+7.3}$ \hunit, reaching a precision on \Ho\ comparable to that achieved by \citet{Birrer20} with seven lenses, demonstrating the power of spatially resolved kinematics. However, the limited spatial resolution of the seeing-limited ground-based data meant that the MAD still left a significant uncertainty.  Simulations have shown that with the spatial resolution of the NIRSpec integral field unit (IFU) on board \jwst, one can dramatically improve the precision of the measurement \citep{Yildirim20, Yildirim23, Wang25}.

We present here the first space-based stellar kinematic map for a time-delay gravitational lens system. The data for \lensname\ are from the NIRSpec IFU on \jwst\ and represent a sixfold improvement in angular resolution with respect to the Keck data used by \citet{Shajib23}. Furthermore, the data probe the rest frame \caii\ triplet region, which is believed to be the most accurate probe of stellar kinematics in elliptical galaxies \citep{Barth02}. To achieve maximum precision and accuracy from the data, we modeled the three spectral components (lens galaxy, quasar, and quasar host) simultaneously, accounting for the effects of \jwst's point spread function (PSF). We implemented the improved methodology from \citet{Knabel25} to account for systematic uncertainties arising from the choice of stellar template libraries, as well as other sources, achieving a 1.58\% systematic covariance. A companion paper will present the lensing and dynamical models, along with the related inference of the Hubble constant (Wang et al., in preparation).

The paper is organized as follows. In \secref{sec:observation_data}, we describe the data acquisition and reduction. In \secref{sec:ancillary}, we describe how we obtained additional pieces of information required for the kinematic extraction. In \secref{sec:methods}, we describe the measurement in detail. In \secref{sec:result}, we present the measured kinematic maps and uncertainty estimates, and compare them with the previous measurement of resolved kinematics for the same system. In \secref{sec:conclusion}, we conclude the paper with a discussion and comparison with previous measurements.

\section{Observations and data reduction}
\label{sec:observation_data}

In this section, we first provide a brief description of the lens system \lensname\ in \secref{sec:lens_description}. Then we describe the spectroscopic observation with \jwst-NIRSpec in \secref{sec:kcwi_spectra} and the data reduction procedure in \secref{sec:reduction}.

\begin{figure*}
    \centering
	\includegraphics[width=1\textwidth]{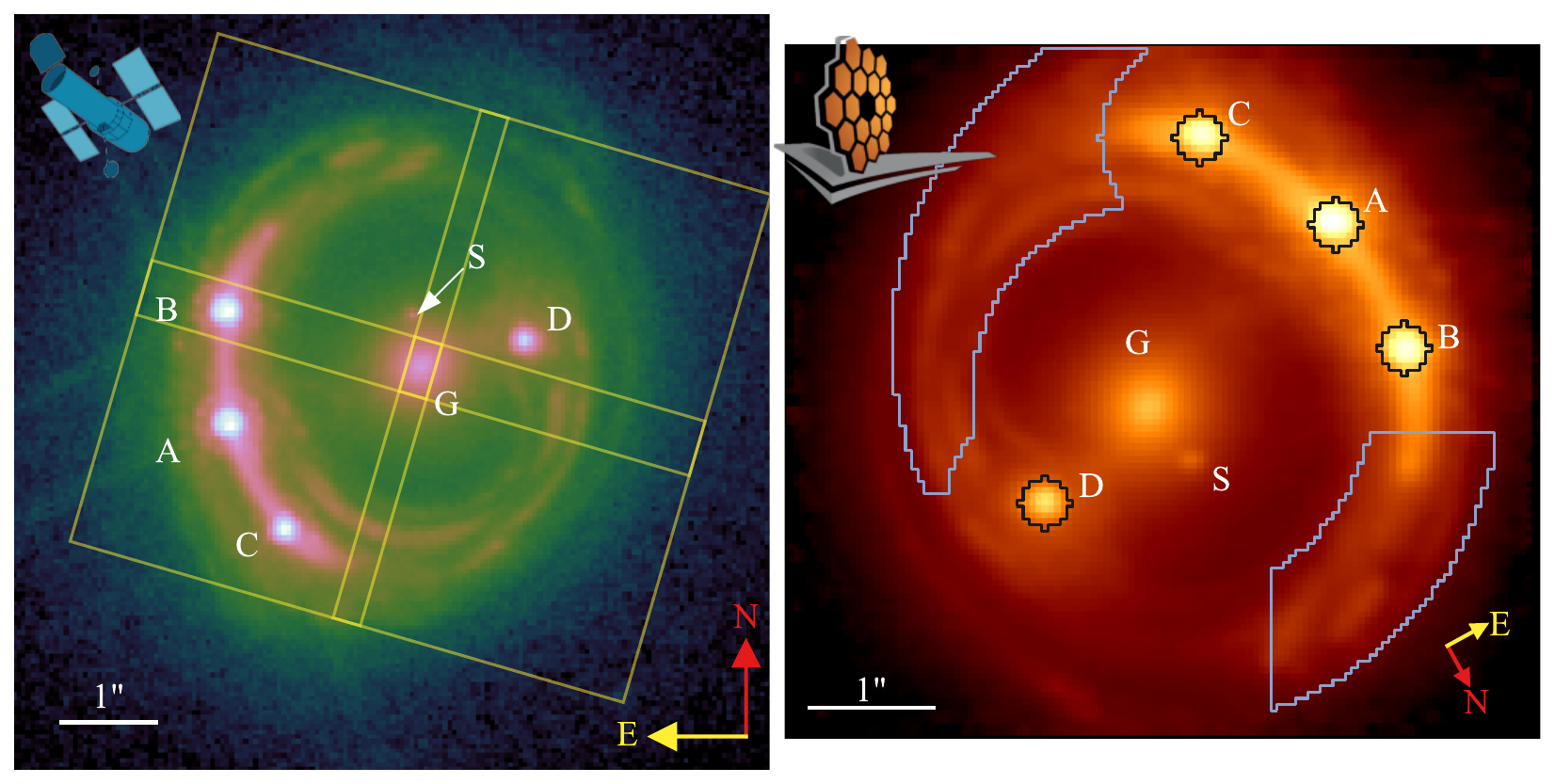}
	\caption{\label{fig:cutout}
	Imaging of the system \lensname. \textbf{Left panel:} HST-ACS image in the F814W band. The four quasar images are labeled A, B, C, and D. The central deflector is marked with G, of which we are measuring the spatially resolved velocity dispersion. An arrow points to the nearby satellite S. The North and East directions, along with a 1\arcsec\ scale, are also illustrated. The $3\farcs2\times3\farcs1$ yellow squares represent the dithered $2\times2$ mosaic pattern of the NIRSpec IFU, which covers a $6\farcs1\times5\farcs55$ field of view over the system. \textbf{Right panel:} "White-light" image of the system from the \jwst-NIRSpec datacube, summed within 8255--8890 \AA\ in the lens galaxy's rest frame. The white bar represents 1\arcsec, and the red and yellow arrows point to the east and north, respectively. The blue contours show the region within which the spaxels are summed to obtain a high-$S/N$ spectrum of the host galaxy. The black regions around the quasar images were used to extract summed spectra for modeling the quasar's emission line components. \textit{HST and JWST logo credits: NASA.}
	}
\end{figure*}

\subsection{Description of the lens system} \label{sec:lens_description}

The quasar lens system \lensname\ was first discovered by \citet{Sluse03}. It comprises an elliptical galaxy as the deflector with a redshift of $z_{\rm d} = 0.295$, while the quasar's (i.e., the lensed source) redshift is $z_{\rm s} = 0.654$ \citep{Sluse07}. Due to the relatively low redshift of the deflector, the system appears brighter and larger in angular size than most other such systems. The intricate features within the Einstein ring offer valuable insights for constraining the lens mass model (see \figref{fig:cutout}).

This system has been extensively studied due to its early discovery and the abundance of lensing information. Time delays for this system were measured by \citet{Tewes13}. \citet{Suyu13, Suyu14} conducted lens modeling and cosmographic analysis of this system. These authors employed simply parameterized lens models based on high-resolution imaging from the \textit{Hubble} Space Telescope's (HST) Advanced Camera for Surveys (ACS; HST-GO 9744; PI: Kochanek), the measured time delays, single-aperture velocity dispersion, and an external convergence estimate to infer $H_0 = 80.0_{-4.7}^{+4.5}$ \hunit. However, as discussed in the introduction, these simply parameterized lens models implicitly break the MSD. The two independent mass profiles used in the analysis by \citet{Suyu14} nonetheless account for additional systematic uncertainties in $H_0$, partly due to the MSD. \citet{Birrer16} independently modeled the mass of this system while accounting for the MSD with a prior on the quasar-host-galaxy size. The findings from these authors revealed that the choice of prior on the dynamical modeling anisotropy dominated the systematic uncertainties in inferring \Ho. These obstacles are overcome by the new data presented in this work.

\subsection{\jwst-NIRSpec spectroscopy} \label{sec:kcwi_spectra}

The \jwst-NIRSpec IFS for \lensname\ was obtained through Cycle 1 program \jwst-GO-1794 (PI: Suyu, co-PIs: Y{\i}ld{\i}r{\i}m, Treu). The observation was carried out on May 9, 2023. We observed the system with G140M grating and the F100LP filter, which covers the observed wavelength range of 0.97--1.84 $\mu$m with a nominal resolution $R \sim 1000$. We performed a $2 \times 2$ mosaic mapping pattern to obtain a total field of view of $6\farcs1\arcsec \times 5\farcs55$ encompassing the entire lensing system, including the lensed quasar images and the arcs from the quasar host galaxy. For individual tiles in the mosaic, we performed a four-point dithering with sub-arcsecond shifts. We used the "IRS2" readout mode with the "NRSIRS2" readout pattern to achieve smaller noise levels for the outer regions with lower surface brightness of the lensing galaxy. For each of the dither positions, we exposed for 1050.4 s with two integrations and seven groups to prevent saturation of the pixels from the bright quasar images. For each mosaic tile, we also obtained a multi-shutter array (MSA) leakage calibration exposure of 1050.4 s. Additionally, we obtained a background exposure from a nearby empty patch of the sky with an exposure time of 1050.4 s to mitigate potential contamination from stray light.

\subsection{Data reduction} 
\label{sec:reduction}

We developed a custom data reduction pipeline \textsc{RegalJumper},\footnote{\url{https://github.com/ajshajib/regaljumper}}, which is built on the standard JWST data reduction pipeline.\footnote{\url{https://jwst-pipeline.readthedocs.io}; we used version 1.18.0, with \textsc{stcal} version 1.12.0 and \jwst\ Calibration References Data System context 1364.} Our custom pipeline includes additional steps to clean up $1/f$ noise, cosmic rays (CRs), and artifacts, in addition to the standard three-stage data reduction procedure implemented by the standard pipeline. Below, we describe how these additional steps are interleaved within the standard reduction stages.

\begin{itemize}
	\item {In Stage 1}, the standard pipeline performs detector-level corrections to generate countrate images (i.e., \texttt{rate} files) per exposure. We included the standard steps: group scale correction, data quality initialization, saturation detection, superbias subtraction, reference pixel correction, linearity correction, dark current subtraction, jump detection with default settings for CR shower flagging, ramp fitting, and gain scale correction. After these steps, we manually identified outliers in the produced files, which are most likely to be unidentified hot pixels, and marked them, along with their adjacent pixels, as bad pixels. We then cleaned the $1/f$ noise using the software package \textsc{NSClean} \citep{Rauscher24}. In this step, we used a hand-drawn trace mask\footnote{included in the GitHub repository of {\sc RegalJumper}} to achieve a robust cleaning, as recommended by \citet{Rauscher24}. We also added "snowball" regions detected previously in this stage to this mask for enhanced robustness. Next, we performed additional CR cleaning using the Python package \textsc{lacosmic} \citep{vanDokkum01}. Next, we subtracted the median of the obtained "leakcal" exposures to manually perform the MSA leakage correction.
	\item {In Stage 2}, the standard pipeline performs additional instrument-level and observing-mode corrections to produce flux-calibrated exposure files (i.e., \texttt{cal} files). We included the standard steps: assigning the world coordinate system (WCS),  MSA flag-open correction, source type determination, flat-field correction, path-loss correction, and photometric calibration. After these steps, we manually detected and flagged outliers and their adjacent pixels again, which are likely to be unidentified hot pixels in the flat field reference frame.
	\item {In stage 3}, the standard pipeline combines the individual exposures to construct the 3D datacube. We used the drizzling method to build the cube with a pixel scale of 0\farcs05 and kept the coordinate system aligned with the IFU frame. After that, we manually performed background subtraction, following the procedure recommended by the TEMPLATES team\footnote{\url{https://github.com/JWST-Templates/Notebooks/blob/main/nirspec_ifu_cookbook.ipynb}} to subtract a simulated background level using the JWST backgrounds tool (\textsc{jbt}).\footnote{\url{https://jwst-docs.stsci.edu/jwst-other-tools/jwst-backgrounds-tool}} Although we did not use the dedicated background exposures to subtract the mean background level, we constructed the background datacube following the same reduction procedure and used it to estimate the background noise level to add to the total noise level of the data during further analysis. The reduced spectra have a reciprocal dispersion of 6.36 \AA\ per pixel.
\end{itemize} 

After constructing the datacube, we correct the "wiggles" in the spectra, which are a manifestation of the resampling noise due to undersampling of the PSF \citep{Law23, Perna23}. We used the software package \textsc{raccoon}\footnote{\url{https://github.com/ajshajib/raccoon}} \shajibraccoon\ to clean the wiggles from the spectra. We describe the settings we used for this step in \appref{app:wiggle_correction} and illustrate an example of the wiggle-cleaned spectra.

\section{Ancillary ingredients for kinematic extraction}
\label{sec:ancillary}

To achieve an accurate absolute calibration of the stellar velocity dispersion, we measured the effective line spread function, as described in \secref{sec:lsf}. We then constructed a lens model of the white-light image obtained from the data cube to properly reconstruct the effective PSF and account for cross-contamination by the multiple spectral components (\secref{sec:lens_modeling}).

\subsection{Line spread function}
\label{sec:lsf}

\shajiblsft\ provide the wavelength-dependent formula for the instrumental dispersion $\sigma_{\rm inst}$ assuming a Gaussian line spread function (LSF) as
\begin{equation}
	\sigma_{\rm inst} (\lambda) = \left[\frac{{\sigma^\prime_{\rm piv}}^2}{\{1 + \alpha \, (\lambda - \lambda_{\rm piv}) \, / \, [{\rm \mu m}]\}^2} - \sigma_{\rm PN}^2 \right]^{1/2},
\end{equation}
where $\sigma^\prime_{\rm piv} = 117.01 \pm 2.25$ \kmps, $\alpha=0.668 \pm 0.035$, $\sigma_{\rm PN} = 6.90 \pm 0.49$ \kmps, and fixed $\lambda_{\rm piv} = 13539.59$ \AA. {This LSF was derived by fitting narrow emission lines from the planetary nebula SMP LMC 58 observed with the same grating and filter configuration of NIRSpec (Program \jwst-CAL-1492, PI: T.~Beck).} At the observed wavelength of the \caii\ triplets (corresponding to rest-frame mean wavelength of 8570 \AA), the instrumental dispersions are $140.6 \pm 3.0$ \kmps\ and $113.0 \pm 2.0$ \kmps\ for the lens and the quasar-host galaxies, respectively. The corresponding LSF full widths at half maximum (FWHMs) are $12.16 \pm 0.26$ \AA\ and $12.48 \pm 0.22$ \AA, respectively.

\subsection{Lens modeling of the white-light image} \label{sec:lens_modeling}

\begin{figure*}
	\includegraphics[width=\textwidth]{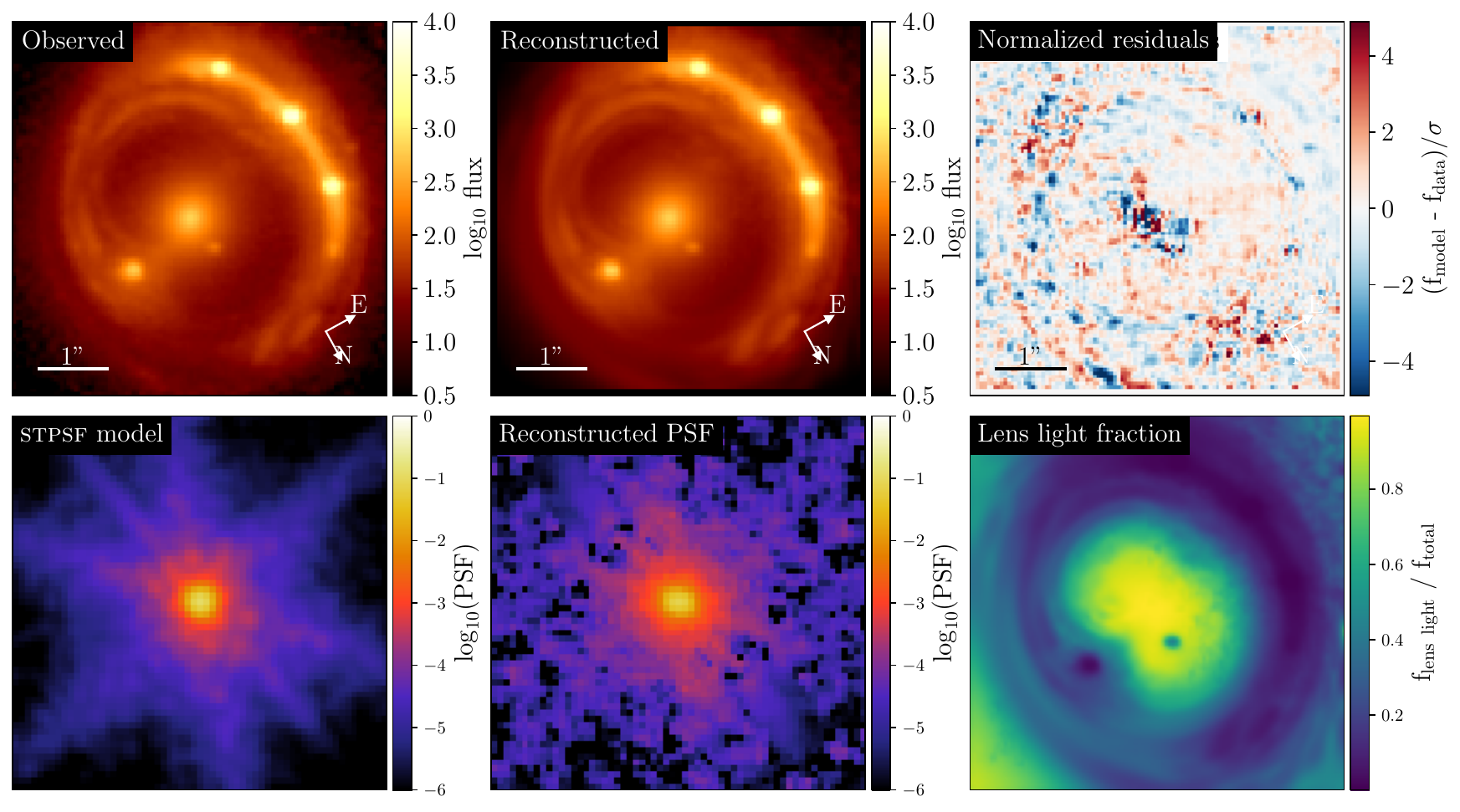}
	\caption{\label{fig:lens_model}
	Lens modeling with the image obtained by summing the datacube within 8700--8800 \AA\ in the lens galaxy's rest frame. This wavelength range covers the Ca triplets of the lens galaxy. \textbf{Top row}, from left to right: Illustration of the data, the optimized-model-based reconstruction of the data, and the normalized residuals.
    \textbf{Bottom row}, from left to right: Illustration of the initial PSF model from \textsc{stpsf}, the iteratively reconstructed PSF, and the fraction of the lens light compared to the total light in the image. The reconstructed PSF will be necessary for dynamical modeling, and we use the lens light fraction to obtain the $S/N$ of the lens galaxy for Voronoi binning.}
\end{figure*}

We performed lens modeling of the image shown in \figref{fig:lens_model}, which was obtained by summing the datacube between 8700 and 8800 \AA\ in the lens rest frame. This wavelength range is immediately redward of the \caii\ region, and we chose this range to obtain the lens galaxy's light profile close to the \cat\ to deblend the lens galaxy's light fraction relative to the other components, thereby accurately extracting the lens galaxy's $S/N$ for use in Voronoi binning. Additionally, the constrained lens galaxy light profile would be used as the tracer distribution in the Jeans dynamical modeling, along with the effective PSF at that wavelength that is reconstructed simultaneously with the lens modeling. From this lens model, we obtained the lens galaxy's light profile for estimating the signal-to-noise ratio ($S/N$) at each spaxel. During lens modeling, we also reconstructed the PSF iteratively, which is necessary for dynamical modeling with kinematic maps. We added the background noise level estimated from the background exposure's datacube to the noise level provided by the reduction pipeline, which is a combination of the propagated detector-level Poisson noise, the read noise, and the flat field's uncertainty levels.

We performed lens modeling with the publicly available software package \textsc{lenstronomy} \citep{Birrer18, Birrer21a}. For the mass model, we used an elliptical power-law (EPL) model \citep{Tessore15} and a residual shear field (also called external shear). We modeled the satellite galaxy's mass distribution with a singular isothermal sphere with a fixed Einstein radius $\theta_{\rm E} = 0\farcs2$, which is the best-fit value from \citet{Suyu13}. We modeled the light distribution of the lens galaxy as a superposition of two concentric elliptical S\'ersic profiles \citep{Sersic68}. We also modeled the satellite's light profile with a circular S\'ersic profile, with fixed $R_{\rm eff} = 0\farcs01$ and $n_{\textrm{S\'ersic}} = 1$ following \citet{Suyu13}. We modeled the quasar-host galaxy's light distribution with a basis set including one elliptical S\'ersic profile and a set of exponential shapelets with $n_{\rm max}=20$ \citep{Refregier03, Birrer15, Berge19}, with their centroids joined. We modeled the quasar point images with the PSF, with their positions freely varied on the image plane.

We obtained an initial PSF estimate using the \textsc{stpsf} software package\footnote{\url{https://github.com/spacetelescope/stpsf}} \citep{Perrin14}, which is a mean of the generated PSFs at six uniformly spaced wavelengths within the wavelength range 8700--8800 \AA, that was used to construct the white-light image. We then iteratively reconstructed the PSF along with the lens model optimization \citep{Chen16, Shajib19}. The reconstructed image is shown in \figref{fig:lens_model}. We used \textsc{Astropy}\footnote{\url{https://www.astropy.org}} \citep{AstropyCollaboration22} to fit a 2D elliptical Gaussian profile to the reconstructed PSF. We find agreement with previous results \citep[e.g.,][]{DEugenio24} that the NIRSpec PSF is slightly more elongated along the $x$-axis (FWHM 0\farcs1666) than along the $y$-axis (FWHM 0\farcs1308). The geometric-mean or circularized FWHM of the PSF is 0\farcs1476.

\section{Extraction of the kinematic maps}
\label{sec:methods}

In this section, we describe our methodology to extract the resolved stellar kinematics from the \jwst-NIRSpec datacube. We generally follow the methodology presented by \citet{Knabel25} to account for the systematic uncertainties stemming from the choice of stellar template libraries, in addition to other sources of systematic uncertainties. For this purpose, we developed the \textsc{squirrel} pipeline,\footnote{\url{https://github.com/ajshajib/squirrel}} which is built on the penalized PiXel Fitting (\textsc{pPXF}) software package\footnote{\url{https://pypi.org/project/ppxf}} \citep{Cappellari17, Cappellari23}. We describe several individual steps in our kinematic fitting procedures in Sects.~\ref{sec:voronoi_bins}--\ref{sec:systematic} and then summarize the steps involved in the entire fitting procedure in \secref{sec:procedure_summary}.

\subsection{Voronoi binning} \label{sec:voronoi_bins}

For appropriate Voronoi binning \citep{Cappellari03}, we first set a target $S/N$ for each bin. We define a "specific" $S/N$ (s$S/N$) as $\textrm{s$S/N$} \equiv (S/N) / \sqrt{\Delta L}$, where $S$ is the summed flux within a wavelength range $\Delta L$ and $N$ is the noise summed in quadrature within the same range. The $\sqrt{\Delta L}$ term standardizes the s$S/N$ as it cancels out the improvement in the $S/N$ solely due to an increase in the summed wavelength range. We took the wavelength range 8700--8830 \AA\ in the lens rest frame, slightly redward of the \caii\ triplets, to compute the s$S/N$.
We set our target s$S/N$ for each Voronoi bin to 90 $\AA^{-1/2}$, which we determined through an trial-and-error experiment to be sufficient for providing stable kinematic measurements (i.e., devoid of extremely high or low spurious values that are indicative of amplified systematic effects due to suboptimal $S/N$).\footnote{For comparison, the equivalent $S/N$ of s$S/N$ $=90\ \AA^{-1/2}$ in the $\AA^{-1}$ unit would be $\sim$40 $\AA^{-1}$, with the conversion peformed through $S/N$-per-pixel over pixel size. The equivalent $S/N$ per logarithmically rebinned spectral pixel of 70 \kmps\ is $\sim$128, to enable comparisons based on the $S/N$ definition adopted by the Sloan Digital Sky Survey \citep{Bolton12} and the MaNGA survey \citep{Westfall19}.}

\begin{figure*}
	\includegraphics[width=\textwidth]{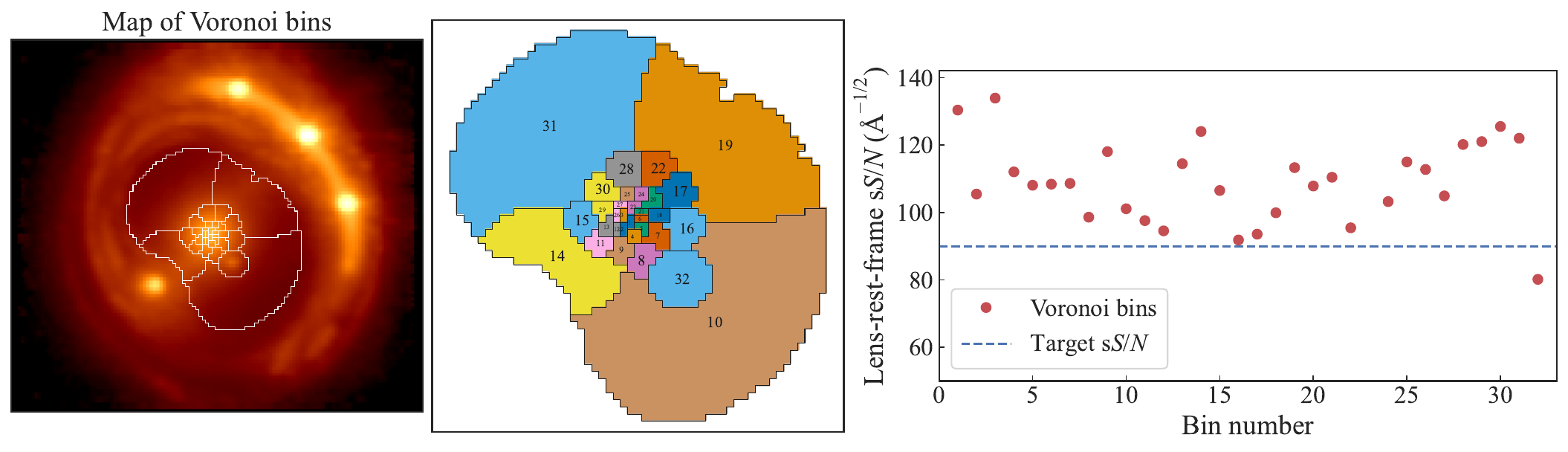}
	\caption{\label{fig:voronoi_binning}
	Voronoi binning map. \textbf{Left panel:} Voronoi bins outlined over the white-light image of the system. \textbf{Middle panel}: Bin numbering. \textbf{Right panel}: Total s$S/N$ for each Voronoi bin. The horizontal dashed line marks 90 $\AA^{-1/2}$, which we set as the minimum target for each bin. Only the last Voronoi bin, which contains the satellite galaxy, is slightly below this s$S/N$ target.
	}
\end{figure*}

We estimated the s$S/N$ of the lens galaxy spectrum for each spaxel by considering the fractional contribution of the lens galaxy's light, as determined by the best-fit model described in \secref{sec:lens_modeling}. This involved scaling the total s$S/N$, which includes signals from all components, by the lens galaxy's light fraction at each spaxel (see \figref{fig:lens_model}). For Voronoi binning, we included only those spaxels that are within 1\farcs5 of the lens galaxy center and have more than half of the total flux contributed by the lens galaxy. We also manually exclude from the initial Voronoi binning an aperture around the satellite galaxy, where the satellite's light is more than 5\% of the lens galaxy's light. We used the software package\footnote{\url{https://pypi.org/project/vorbin}} \textsc{vorbin} to perform Voronoi-binning with the minimum target s$S/N$ per bin. We then manually combined a few single-spaxel bins near the center so that a Voronoi bin includes at least two spaxels and a few Voronoi bins in the outer periphery for which the initial Voronoi binning did not lead to a total s$S/N$ $\geq 90\ \AA^{-1/2}$. Finally, we included back the aperture containing the satellite galaxy as the last bin, yielding a total of 32 Voronoi bins. The Voronoi binning map and the s$S/N$ in each bin are illustrated in  \figref{fig:voronoi_binning}.

\subsection{Log-rebinning of spectra} \label{sec:rebinning}

We rebinned the spectral data, which is linearly sampled in wavelength, into a logarithmically sampled one using \textsc{pPXF}. Such log-rebinning is a standard procedure for kinematic measurements, facilitating numerical fitting with a constant velocity scale per wavelength-pixel \citep{Cappellari17}. We chose the default rebinning settings of \textsc{pPXF} that maintain the same pixel numbers before and after rebinning, resulting in a velocity scale of 171.92 \kmps. We also created a covariance matrix for the log-rebinned spectra, as log-rebinning induces covariances within the spectra, with stronger covariances between more closely adjacent bins. We created the covariance matrix using the Monte Carlo method, by generating 5\,000 realizations of the spectra that are statistically consistent with the data within its noise levels and then taking the covariance of these samples after log-rebinning them. This covariance enabled us to fully propagate the uncertainty provided by the reduction pipeline into the final analysis by providing it to the \textsc{pPXF} kinematic fitting routine.

\subsection{Spectral templates} \label{sec:templates}

We chose two stellar template libraries: the Indo-US \citep{Valdes04} and the X-shooter Spectral Library (XSL) DR3 \citep[][]{Verro22}. We used the "cleaned" libraries provided by \citet{Knabel25}.\footnote{The "cleaned" libraries can be obtained from \url{https://github.com/TDCOSMO/KINEMATICS_METHODS}.} We did not use the MILES library, which was also cleaned by \citet{Knabel25}, since the original MILES library does not cover wavelengths up to the \cat. Although the \cat\ library from \citet{Cenarro01} is referred to as the MILES-CaT library, because it contains a subset of the MILES stars, we did not use this library as it was not cleaned by \citet{Knabel25}.

From these libraries, we further removed spectra that have gaps within the fit wavelength ranges around the lens and quasar-host galaxy's \caii\ triplet regions, either by using the flags provided \citet{Knabel25} or by directly checking for such gaps within the relevant wavelength range. Ultimately, we had 609 stellar templates in the Indo-US library and 489 stellar templates in the XSL. The Indo-US library has a constant resolution FWHM of 1.35 \AA\ \citep{Beifiori11}, which corresponds to instrumental dispersion (standard deviation) $\sigma_{\rm temp} = 18$ \kmps. The XSL has an instrumental dispersion $\sigma_{\rm temp} = 11$ \kmps\ (for the VIS arm), corresponding to a resolution FWHM of 0.74 \AA\ at the wavelength of the \caii\ triplet. {The stellar template libraries are broadened by an additional $(\sigma_{\rm inst}^2 - \sigma_{\rm temp}^2)^{1/2}$ before being used in the kinematic fitting to account for the instrumental broadening in the observed absorption line widths.}

The \caii\ triplet region of the lens galaxy coincides with the H$\alpha$, [N \textsc{ii}], and [S \textsc{ii}] lines from the quasar and the host galaxy. Therefore, we fit the spectra with multiple components {\citep[similar to][]{Turner24}}: the lens galaxy's stellar continuum and absorption features, the host galaxy's stellar continuum and gas lines, and the quasar emission lines. Below, we describe how we built a set of optimal templates for each of these components.

\subsubsection{Template for the host galaxy's stellar continuum} \label{sec:host_template}

\begin{figure}
	\includegraphics[width=0.5\textwidth]{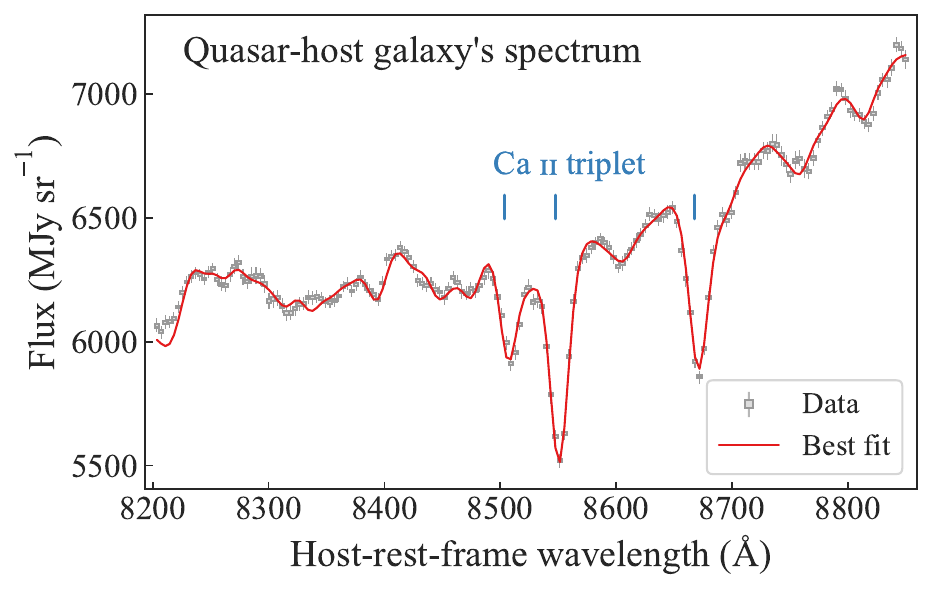}
	\caption{\label{fig:host_cat_fit}
	Fitting of the \caii\ region of the quasar host galaxy. The gray bars represent the data points, with the width of each bar corresponding to the pixel size and the height representing the original $\pm 1\sigma$ noise level. The gray lines attached to the bars represent the boosted uncertainty levels to achieve $\chi^2_{\rm red} = 1$. The red line illustrates the best-fit model. The \cat\ features are marked with blue lines.
	}
\end{figure}

To create a template for the quasar-host galaxy's stellar continuum, we first extracted a summed spectrum within 8200--8850 \AA\ in the host galaxy's rest frame from a region containing the lensed arcs (shown with blue contours in the right panel of \figref{fig:cutout}). Then, the spectra were fit with the Indo-US and XSL template libraries, while also accounting for the contribution from the lens galaxy's stellar continuum at the corresponding wavelength range. This capability of \textsc{pPXF} to fit multiple components with different kinematics is crucial for modeling the complex spectrum analyzed in this paper, where emission lines from both the lens galaxy and the quasar plus host (at significantly different redshifts) can have multiple kinematic components (e.g., broad or narrow profiles or different velocities). We followed the methodology of \citet{Knabel25} to select the combination of an additive polynomial degree of four and a multiplicative polynomial degree of one that yields a subpercent difference in the best-fit velocity dispersion between the Indo-US library and XSL, while orders any higher do not significantly affect the results. \figref{fig:host_cat_fit} shows the best-fit model to the host galaxy's \caii\ triplet.

Using the above fit, we constructed a single template for the host galaxy's stars by combining the Indo-US library with the corresponding best-fit weights. The wavelength range around the lens galaxy's \cat\, which is fit later in this analysis, corresponds to 6463--6960 \AA\ in the quasar-host galaxy's rest frame. The Indo-US stellar library, and hence the single template constructed above, has coverage down to this wavelength range. The quasar-host galaxy does not have any prominent kinematic lines around the lens galaxy's \cat, and thus any wavelength-dependent difference between the quasar-host galaxy's \cat\ wavelength and that corresponding to the range around the lens galaxy's \cat\ would have a negligible impact on our analysis. Furthermore, although we fit the host galaxy spectra summed from a region primarily containing the disk, rather than the bulge, which may have different kinematic and stellar population properties, this difference is unlikely to affect our analysis for the same reasons mentioned above.

\subsubsection{Templates for the quasar spectra} \label{sec:quasar_template}

\begin{figure}
	\includegraphics[width=\columnwidth]{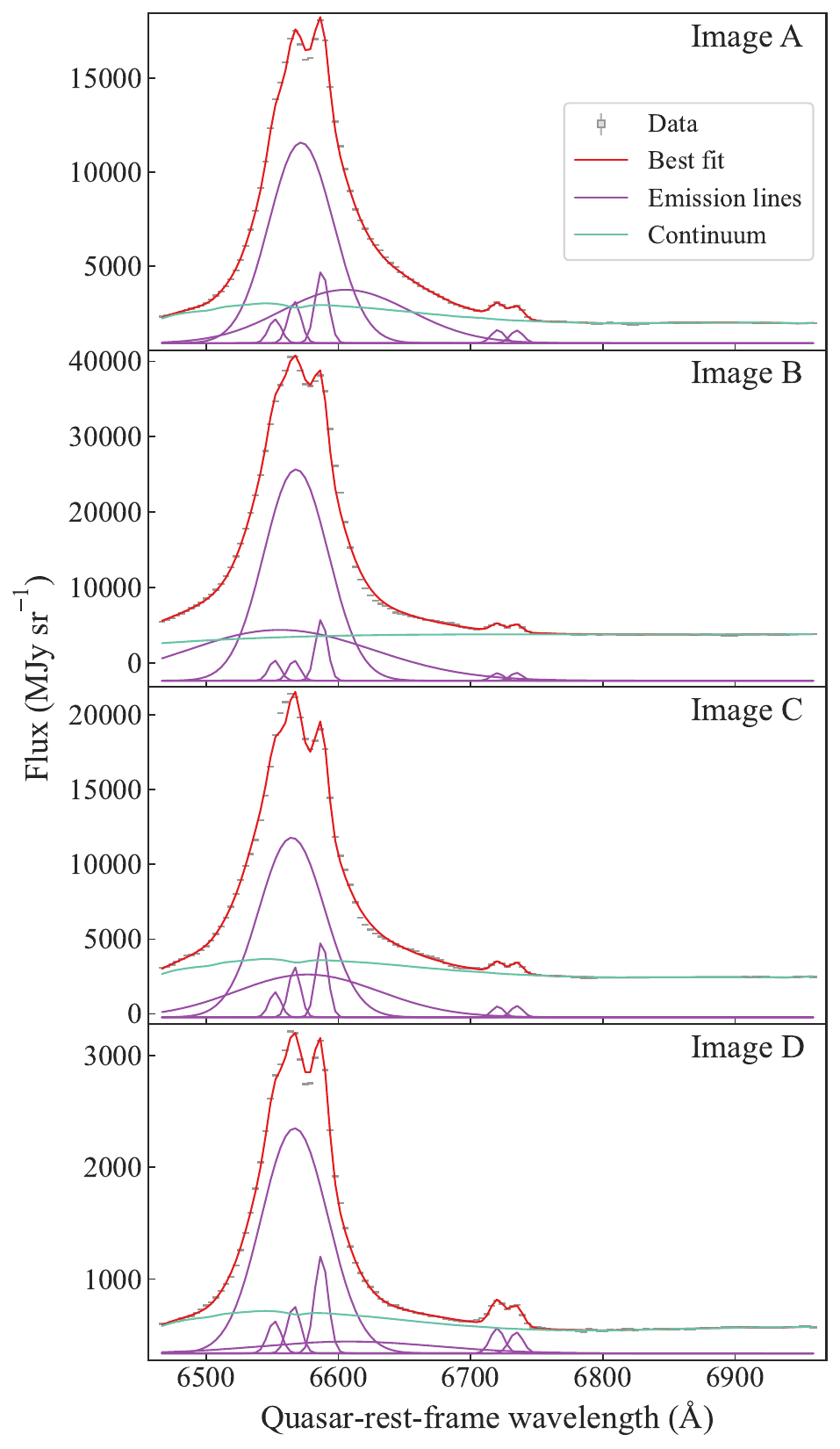}
	\caption{\label{fig:quasar_fit}
	Fitting of the quasar lines at images A, B, C, and D, respectively, from top to bottom. The observed spectra are shown with gray bars, with the width corresponding to the bin width and the height representing the $\pm1\sigma$ noise range. The red lines show the best fit. The individual narrow and emission lines from the best fit are shown in purple, and the lens galaxy's stellar continuum is shown in emerald. We adopted the best-fit gas lines with fixed relative amplitudes for all three images as free linear components in our kinematic fitting of the spectra across the entire field of view.
	}
\end{figure}

To build the templates for the quasar images, we extracted summed spectra from circular apertures of 0\farcs2 radius around the four quasar images (illustrated in \figref{fig:cutout}). To each of these spectra, we fit multiple emission lines: narrow [N \textsc{ii}] and [S \textsc{ii}] doublets, and H$\alpha$ with both narrow and broad components. While each narrow line was fit with a single Gaussian profile with a common velocity shift and dispersion for all of them, the broad H$\alpha$ line was fit with two Gaussians with independent velocity shifts and dispersions, following \citet{Guo18}.\footnote{For example, see \url{https://nbviewer.org/github/legolason/PyQSOFit/blob/master/example/example.ipynb}.} Here, we aimed to construct the templates for the quasar images' emission lines independently of the physical interpretation of the components. We used an additive polynomial of order four and a multiplicative polynomial of order one to account for the quasar's continuum.
Additionally, the host galaxy's combined stellar template from \secref{sec:host_template} was also incorporated as an additional component in the fit, with its associated best-fit velocity and dispersion kept fixed as obtained in that section. We combined the best-fit emission lines for a given quasar image to construct its associated template (\figref{fig:quasar_fit}). In our fitting of the lens galaxy spectra, we used all four templates as independent components because they have different profile shapes, likely due to microlensing \citep[e.g.,][]{Hutsemekers24}, and a given spaxel can receive contributions at varying levels from the four quasar images carried through the PSF spikes. The constructed quasar templates excluded the quasar continuum because it would be accounted for by the polynomials adopted in the kinematic fitting of the lens galaxy's spectra. {We illustrate the expected and effective levels of quasar contribution in each Voronoi-binned spectrum in \appref{app:quasar_contamination}.}

\subsubsection{Templates for the host galaxy's emission lines}

We also construct a set of emission line templates for the narrow H$\alpha$, [N \textsc{ii}], and [S \textsc{ii}] lines from the host galaxy, which are independent of the quasar's narrow lines described above in \secref{sec:quasar_template}. If these emission line components were allowed to have individually free amplitudes, they would sometimes result in unrealistic line ratios, especially in cases where their fractional contribution to the spectrum is relatively small. Therefore, we built eight emission line templates where the line ratios between them were fixed within each template. The number eight was adopted because the NIRSpec PSF is expected to have eight prominent spikes; thus, a given spaxel can receive contributions from eight different positions of the lensed arc in a nontrivial manner. We divided the regions traced with blue contours in \figref{fig:cutout} into eight segments and fit the spectra summed within each of these segments. In addition to the emission lines with freely varying amplitudes, we included all the other components in these fits, namely the quasar templates, the host galaxy's stellar continuum template, and the Indo-US template library for the lens galaxy's spectra. We then constructed the emission line templates from each of the fits by fixing the relative line ratios to the best-fit values. When fitting the lens galaxy spectra in subsequent steps, including these eight emission line templates, each template was allowed to have independent kinematic properties (both in mean velocity and dispersion) because kinematic properties from different points of the quasar-host galaxy can mix in a single spaxel due to nontrivial superposition of the PSF spikes, as mentioned above. The mean velocity of the host galaxy's \cat\ lines obtained in \secref{sec:host_template} is $293 \pm 7$ \kmps, and the mean velocity of the emission lines obtained in this section is $217.0 \pm 0.4$ \kmps. This difference suggests potentially distinct kinematic properties of the stars and gas that generated the fit absorption and emission features, respectively. Interpreting these velocities as residual systemic velocity would lead to adjusting the redshift of the host galaxy (and its quasar) by approximately $+0.001$ to $z_{\rm s} \approx 0.655$.

\subsubsection{Template for the lens galaxy's spectra} \label{sec:lens_template}

We constructed the stellar template for the lens galaxy in two regions: one for the Voronoi bins with luminosity-weighted centers within a 0\farcs5 radius from the lens galaxy center, and the other for those outside (for bins 10, 14, 19, and 31). We excluded bin 32, which contains the satellite galaxy, from this step in the template-making process. For a given set of settings on the polynomial orders and the fit wavelength range, the summed spectra from these two regions are fit with a given template library from the Indo-US and XSL, and the template library was combined using the best-fit weights to produce one single template associated with the given settings and template library. We used this combined template when fitting to the spectra for all the Voronoi bins within the associated region to mitigate spurious scatter in the fit kinematics, following standard procedure in the literature \citep{Shajib23, Knabel24}.

\subsection{Fitting spurious spikes}

Some spurious spike-like features in the spectra are not well-fit by the stellar templates, and they also do not align with any expected emission lines at the redshifts of the lens or quasar-host galaxy. We explicitly modeled these spurious spikes as emission lines with instrumental resolution, since otherwise, they can have a significant impact on the measured kinematics. These spurious spikes may have astrophysical origins, such as emission lines from faint sources at other redshifts with undetected continua or the Zodiac, or may be data-level defects, such as artifacts from unidentified warm pixels or uncleaned CRs. For a given spectrum and a set of fitting ingredients and settings, we first identified the 5$\sigma$ outliers or spurious features. Then, we refit the spectra with the same ingredients but now with the identified spikes modeled. Next, we identified 3$\sigma$ outlier pixels and refit the spectra again with all the identified spikes modeled. Finally, we identified 2.7$\sigma$ outlier pixels. We performed this identification process in multiple steps until we reached our outlier detection threshold of 2.7$\sigma$ for a more robust detection of the true outlier at our target threshold. Furthermore, we performed the same detection procedure for a given Voronoi-binned spectrum using both template libraries and selected only the spikes identified in both cases. We set our outlier detection threshold at 2.7$\sigma$ because the statistically expected number of pixels with absolute residuals larger than 2.7$\sigma$ is less than one, given the total number of pixels within the fit wavelength range. Although most of the identified spikes have positive amplitudes, a fraction of them have negative amplitudes, and we allowed for this in our model.

\subsection{Estimating systematic uncertainty} \label{sec:systematic}

We accounted for the systematic uncertainties stemming from several aspects, where a particular choice needs to be made for the kinematic fitting, namely the template stellar library, the orders of the additive and multiplicative polynomials used to model the continuum, the detection threshold for spurious spikes, and the fit wavelength range. We adopted multiple viable choices for each of these aspects to marginalize over them and account for the associated systematic covariances. We took the following choices for these aspects:
\begin{itemize}
	\item Template library: Indo-US and XSL;
	\item Order of additive and multiplicative polynomials: (2, 2), (3, 2), (4, 2), (2, 3);
	\item Spurious spike detection threshold: 2.7$\sigma$ and 2.6$\sigma$; and
	\item Fit wavelength range: 8255--8890 \AA, 8280--8890 \AA, and 8255--8865 \AA.
\end{itemize}
By combining these choices, we had a total of 48 model setups. For each choice, we regenerated combined templates for the two regions described in \secref{sec:lens_template} to fit the Voronoi-binned spectrum. We first combine the {16} model setups within each given fit wavelength range choice using the Bayesian information criterion (BIC) weighting following \citet{Knabel25}. Then, we combined the three sets of kinematic measurements and their covariances (mean velocities and velocity dispersions) with equal weights. We did not use BIC weighting in this latter case, as it would weigh more heavily the case with a smaller number of data points, which is undesirable.

\subsection{Kinematic fitting procedure} \label{sec:procedure_summary}

Here, we summarize the steps involved in our fitting procedure, which combines all the ingredients and individual steps described above.

\begin{enumerate}
	\item We performed Voronoi binning as described in \secref{sec:voronoi_bins}.
	\item We log-rebinned the Voronoi-binned spectra and produced the covariance matrix of the rebinned spectra propagated from the detector-level uncertainties as described in \secref{sec:rebinning}. {We used the full covariance matrix as input to \textsc{pPXF}, a first for kinematic measurements in the literature.}
	\item We prepared the stellar templates for both the lens and quasar-host galaxies, the quasar templates, and the emission line templates (associated with the host galaxy) described in \secref{sec:templates}. For each of the 48 model combinations described in  \secref{sec:systematic}, we generated the combined templates within two annuli: one that includes all the Voronoi bins with luminosity-weighted centers within 0\farcs5 of the center, and the other including the rest (but excluding bin 32 since it is contaminated by the satellite galaxy).
	\item We first boost the covariance of each Voronoi-binned spectrum within the range 8441--8570 \AA, so that the median $S/N$ in this range matches that of the rest. As this wavelength range contains the H$\alpha$ and [N \textsc{ii}] lines from the quasar and its host, which may have a very high $S/N$ in the spaxels they appear brightly in, we adjust the $S/N$ so as not to allow any strong residual in this region to bias the overall fit. We fit this spectrum using the prepared templates and then boosted the covariance of the spectrum by a factor of $\chi^2$ before refitting it to obtain the final kinematic measurements and uncertainties. We treat the formal error on the kinematic measurements provided by \textsc{pPXF} from this final fit as the associated statistical uncertainty. The kinematic values of the host galaxy's stellar component were fixed to the ones obtained from the fit for the host galaxy's \cat\ in \secref{sec:host_template}. However, the narrow emission line sets were allowed to have free mean velocities and dispersions, independently for each set. As a result, the fit included a total of nine kinematic components: the lens galaxy and eight sets of narrow emission line templates.
	\item We combined the models using BIC weighting separately within the three sets with different wavelength ranges, and then combined the resultant three sets of values with equal weighting, as described in \secref{sec:systematic}.
\end{enumerate} 

\begin{figure}
	\includegraphics[width=\columnwidth]{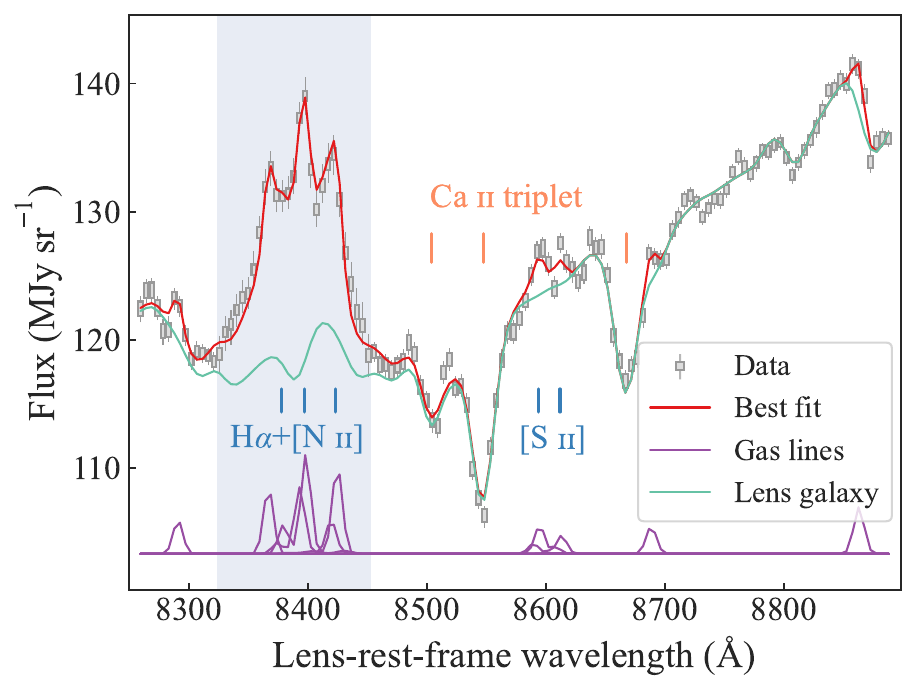}	
	\caption{\label{fig:example_spectra_fit}
	Best fit (red line) of the Voronoi-binned spectra in bin 22 as an example. The gray bars represent the observed data, with the width indicating the bin width and the height representing the original $1\sigma$ noise level. The vertical gray lines attached to the bars represent the boosted $1\sigma$ noise level. These $1\sigma$ noise levels for each pixel are estimated from random sampling given the covariance matrix. The \cat\ wavelengths from the lens galaxy are marked with orange lines, and the wavelengths of the host galaxy's emission lines are marked with blue lines.  The blue-shaded region marks where we boosted the noise levels around the H$\alpha$ and [N \textsc{ii}] lines. The emerald line shows the contribution from the lens galaxy's stars. The sets of narrow emission lines from the quasar host galaxy (including spurious spikes identified close to 8300, 8700, and 8900 \AA\ in this case) are shown in purple. 
}		
\end{figure}

\figref{fig:example_spectra_fit} illustrates the kinematic fitting of bin 22, as an example. The fits for the rest of the bins are illustrated in \appref{app:individual_voronoi_bin_fits}.

\section{Kinematic maps} 
\label{sec:result}

In this section, we provide our measured kinematic maps in \secref{sec:extracted_kinematics} and compare them with previous resolved measurements from ground-based data in \secref{sec:comparison_with_prev}.

\subsection{Kinematic maps from JWST NIRSpec data} \label{sec:extracted_kinematics}

\begin{figure}
	\includegraphics[width=\columnwidth]{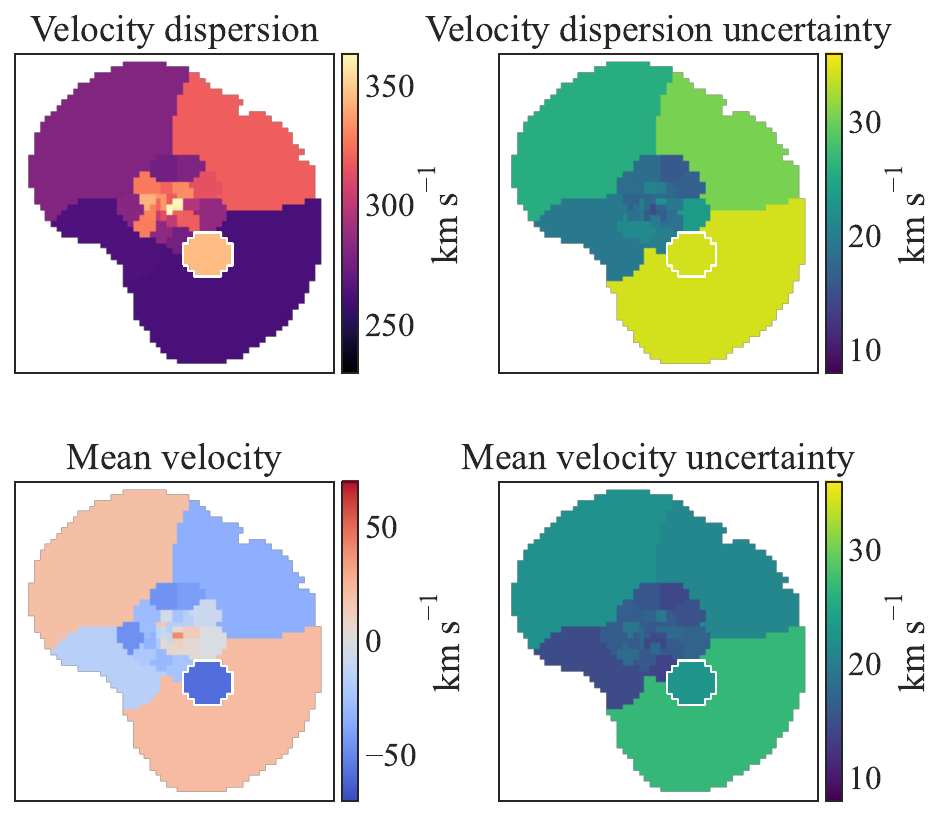}
	\caption{\label{fig:kinematic_maps}
	Velocity dispersion (top row) and mean velocity (bottom row) measurements in 32 Voronoi bins. The first column shows the measurement for each case, and the second column shows the measurement uncertainties. The bin containing the satellite galaxy (i.e., bin 32) is outlined in white. The bins closer to the center have smaller uncertainties, thanks to less contamination from the quasar and its host galaxy. The systemic velocity of 213 \kmps\ was subtracted from the mean velocity map. The mean velocity map indicates that the lens galaxy is a slow rotator, with small mean velocities ($\lesssim$50 \kmps) and a lack of a bipolar velocity profile.
	}
\end{figure}

\begin{figure}
    \centering
	\includegraphics[width=0.9\columnwidth]{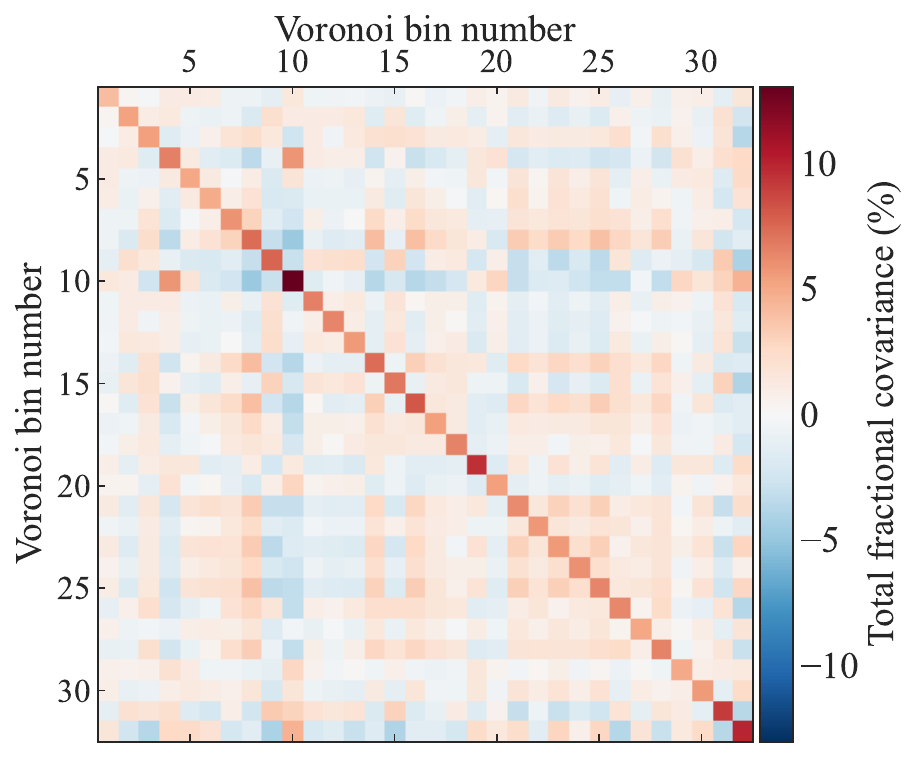}
	\caption{\label{fig:covariance_matrix}
	"Fractional" covariance matrix. In  this matrix, the element with indices $i,j$ is $[{\rm cov} (\sigma_{\rm los, \it i}, \sigma_{\rm los, \it j}) / \sigma_{\rm los, \it i} \sigma_{\rm los, \it j}]^{1/2}$. The diagonal elements represent the combined statistical and systematic uncertainties. The mean for the absolute values of the off-diagonal elements is $1.14\pm0.04$\%.
	}
\end{figure}

The 2D maps of the velocity dispersion and mean velocity measurements in the 32 bins are illustrated in \figref{fig:kinematic_maps}. We show the total covariance matrix of the velocity dispersion measurements in  \figref{fig:covariance_matrix}.
The mean for absolute values of the off-diagonal elements in the "fractional" covariance matrix [i.e., with the $ij$-th element being $\{{\rm cov} (\sigma_{\rm los, \it i}, \sigma_{\rm los, \it j}) / \sigma_{\rm los, \it i} \sigma_{\rm los, \it j}\}^{1/2}$] is $1.24 \pm 0.04$\%. {Although \citet{Yildirim20} forecast a 4\% precision on the \Ho\ measurement based on simulated, 
\jwst-quality 2D kinematics of \lensname, the actual $S/N$ in our data is less than that assumed in their simulations. Therefore, this 4\% precision should be regarded as a lower limit, with the actualized uncertainty on \Ho\ to be presented by Wang et al. (in preparation).}

We find that the lens galaxy is a slow rotator, with mean velocities $\lesssim$50 \kmps\ across the kinematic map. This result agrees with the previous, resolved measurements from the KCWI by \citet{Shajib23}.

In the kinematic map, the bin containing the satellite galaxy (outlined with a white border in \figref{fig:kinematic_maps}) is noticeable for having a slightly different mean velocity and dispersion compared to the surrounding bins. The satellite galaxy was previously attributed as the "satellite" of the main deflector due to its similarity in color \citep{Claeskens06}. Indeed, the satellite galaxy does not exhibit distinctly identifiable spectral features at a separate redshift, suggesting that its redshift is very close to that of the central galaxy. The mean velocity of the bin after subtracting the systemic velocity of the central deflector is $-59 \pm 23$ \kmps. Although this velocity measurement is a weighted combination of both the satellite and main deflector contributions in bin 32, if taken at face value, the $-5\sigma$ value of $-174$ \kmps\  would only change the satellite galaxy's redshift by $5.8\times10^{-4}$, which is below the typical uncertainty levels (i.e., $\sim10^{-3}$) in spectroscopic redshift measurements.

\subsection{Comparison with previous measurement} \label{sec:comparison_with_prev}

\begin{figure*}
	\includegraphics[width=\textwidth]{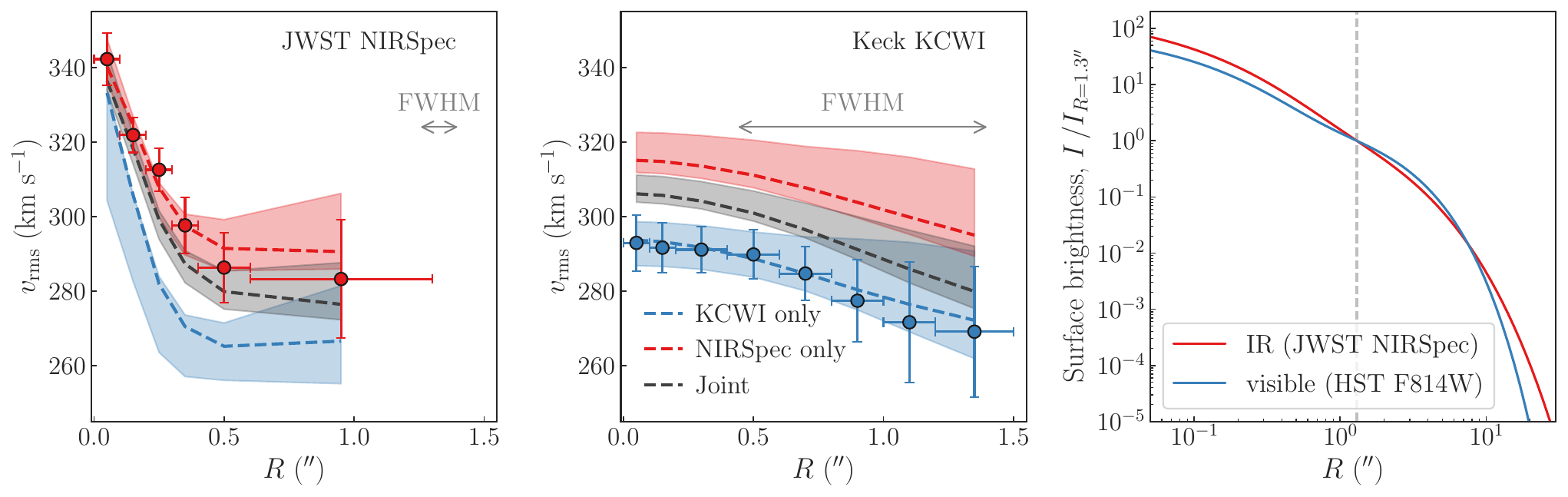}
	\caption{\label{fig:kcwi_compare}
	One-dimensional profile of the rms velocity $v_{\rm rms} \equiv (\sigma_{\rm los}^2 + v_{\rm los}^2)^{1/2}$, where $v_{\rm los}$ is the line-of-sight mean velocity after subtracting the mean systemic velocity of 213 \kmps. \textbf{Left panel:} One-dimensional $v_{\rm rms}$ profile for JWST-NIRSpec measurements from this paper. \textbf{Middle panel:} One-dimensional $v_{\rm rms}$ profile for the Keck-KCWI measurements from \citet{Shajib23}. The vertical error bar represents the total uncertainties, including both statistical and systematic contributions, and the horizontal error bars denote the annulus widths.
	The dashed lines represent the best-fit model predictions from JWST-only data (red), KCWI-only data (blue), and from a joint fit (black), and the associated shaded regions illustrate the 1$\sigma$ extent of the predictions from the model posterior. The horizontal arrows represent the FWHMs of the corresponding PSF in each panel. \textbf{Right panel:} One-dimensional light profile from double S\'ersic fits performed on the NIRSpec white-light image (red line) as part of the lens modeling in this paper, and from that, on the HST imaging in the F814W filter (blue line) from \citet{Shajib23}. The vertical dashed line marks the 1\farcs3 transition radius, where we stitch the light profiles with IR in the inner region and the visible one in the outer region that is used in the dynamical model of the NIRSpec 1D profile. From a likelihood-ratio test, we find that the JWST and KCWI measurements are consistent at $\sim$1.2$\sigma$ (i.e., the likelihood-ratio test statistic's $p$-value is $\sim$0.22).
	}
\end{figure*}

\begin{figure*}
	\centering
	\includegraphics[width=0.6\textwidth]{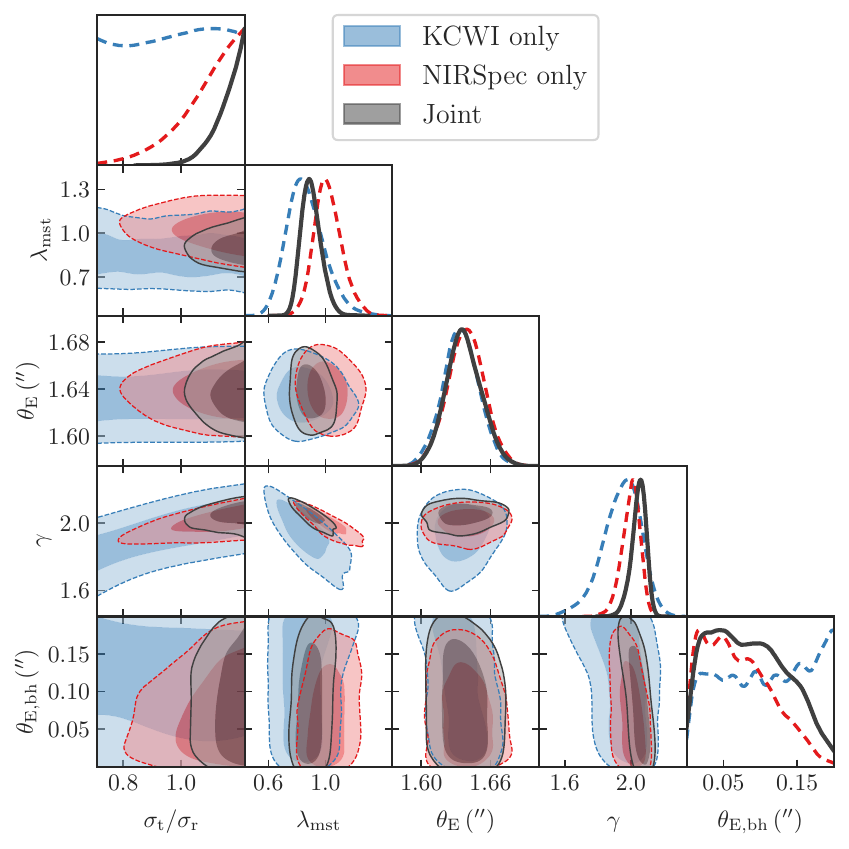}
	\caption{\label{fig:corner}
	Comparison of the parameter posteriors of the dynamical modeling performed with JWST-NIRSpec-only data (red), Keck-KCWI-only data (blue), and both combined (black). The four free parameters in our dynamical model are the ratio of the tangential and radial velocity dispersion $\sigma_{\rm t} / \sigma_{\rm r} \equiv (1 - \beta_{\rm ani})^{1/2}$ with $\beta_{\rm ani}$ being the amplitude of a spatially constant anisotropy profile, the MST parameter $\lambda_{\rm mst}$, the main deflector's Einstein radius $\theta_{\rm E}$, the power-law mass profile's logarithmic slope $\gamma$, and the super-massive black hole's Einstein radius $\theta_{\rm E, bh}$. We have imposed a prior on the total Einstein radius $(\theta_{\rm E}^2 + \theta_{\rm E, bh}^2)^{1/2}$ based on the lens model posterior of \citet{Suyu13}, which dominates the $\theta_{\rm E}$ posterior illustrated here. All of them are consistent at $\lesssim$1$\sigma$ confidence levels. The JWST NIRSpec's superior spatial resolution also allows for the exclusion of a large fraction of the prior volume in the anisotropy parameter ($\sigma_{\rm t}/\sigma_{\rm r}$) that is largely unconstrained by the Keck-KCWI measurements.
	}
\end{figure*}

In this section, we compare resolved kinematic measurements from the JWST-NIRSpec IFS with those from the previous study using the Keck KCWI IFS \citep{Shajib23}. We used the updated values of the KCWI measurements from \citet{TDCOSMO25}, which employed the improved methodology of \citet{Knabel25}. We took the 1D $v_{\rm rms}$ profile as the measurements to compare, where $v_{\rm rms} \equiv (\sigma_{\rm los}^2 + v_{\rm los}^2)^{1/2}$ with the $v_{\rm los}$ being the mean line-of-sight velocity after subtracting off the systemic velocity. We created the 1D $v_{\rm rms}$ profile in annular bins from the 2D kinematics of the NIRSpec data, following the same methodology as \citet{Shajib23}, but with slightly different annulus widths that are suited to the different Voronoi binning map used in this paper (see  \figref{fig:kcwi_compare}). 

In addition to the two datasets having different PSF FWHMs, the tracer populations are slightly different, given that the measurements were taken at different wavelengths. Therefore, we performed the comparison between the two datasets in the model parameter space after performing dynamical modeling of the 1D $v_{\rm rms}$ profiles. For the tracer distribution of the KCWI measurements, we used the double S\'ersic light profile from \citet{Shajib23}, which was fit to a large cutout of the HST F814W imaging after subtracting the lensed quasars and arcs using the lens model from \citet{Suyu13}. In the IR band, imaging data with a larger field of view than that in the NIRSpec data was not available. Therefore, we took the best-fit light profile from the lens modeling performed on the NIRSpec white-light image in  \secref{sec:lens_modeling} and stitched it with the light profile from the HST imaging to cover the outer region. We took 1\farcs3 as the stitching radius since that is the largest extent covered by the 1D $v_{\rm rms}$ profile from the NIRSpec data.

We perform the dynamical modeling of the 1D $v_{\rm rms}$ profile with NIRSpec-only, KCWI-only, and the joint data. We took a spherical power-law mass model and performed Jeans modeling of the kinematics. We took a spherical model for simplicity, given that the kinematic profile to be modeled is 1D. We also added a central super-massive black hole in our mass model, with its mass $M_{\rm bh} \propto \theta_{\rm E, bh}^2$ a free parameter \citep[see][ for its impact on 2D kinematics]{Wang25}. Our model has five free parameters: the Einstein radius $\theta_{\rm E}$, the power-law mass profile's logarithmic slope $\gamma$, the super-massive black hole's Einstein radius $\theta_{\rm E, bh}$, the ratio of the tangential and radial velocity dispersion $\sigma_{\rm t} / \sigma_{\rm r} \equiv (1 - \beta_{\rm ani})^{1/2}$ with $\beta_{\rm ani}$ being the amplitude of a spatially constant anisotropy profile, and the mass-sheet transformation (MST) parameter $\lambda_{\rm mst}$ \citep[for a brief explanation of the MST, see][]{Treu23}. Here, the $\lambda_{\rm mst}$ parameter encapsulates both the contribution from the external convergence (i.e., $\kappa_{\rm ext}$) and the internal MST. We imposed a prior on the total Einstein radius as $(\theta_{\rm E}^2 + \theta_{\rm E, bh}^2)^{1/2} \sim \mathcal{N}(1\farcs640, 0\farcs015)$ based on the lens model of \citet{Suyu13}, but adjusting the mean value given our inclusion of the central black hole with an associated Einstein radius of 0\farcs15. We also imposed a uniform prior on the anisotropy parameter $\sigma_{\rm t} / \sigma_{\rm r} \sim \mathcal{U}\,(0.72, 1.22)$, which imposes the bounds $-0.5 < \beta_{\rm ani} < 0.5$. We performed the dynamical modeling with \textsc{lenstronomy}'s \textsc{Galkin} module and sampled from the posterior using the affine-invariant ensemble sampler \textsc{emcee} \citep{Foreman-Mackey13}. Then, we performed a likelihood ratio test using the statistic $\Lambda \equiv 2 (\log \mathcal{L}_{\rm comb} - \log \mathcal{L}_{\rm NIRSpec} - \log \mathcal{L}_{\rm KCWI}) \approx 7.0$, which is expected to follow a $\chi^2$ distribution with five degrees of freedom. The $p$-value of this $\Lambda$-statistic is 0.22 ($z$-score $\sim$1.2$\sigma$). Therefore, we cannot exclude the null hypothesis that the two datasets are consistent.
It can be noticed in \figref{fig:corner} that the NIRSpec measurements provide more constraint on the stellar anisotropy ($\sigma_{\rm t} / \sigma_{\rm r}$), breaking the MAD more effectively than the previous KCWI measurements with seeing-limited resolution, which left this parameter largely unconstrained within the prior volume.

\section{Conclusion}
\label{sec:conclusion}

In this paper, we extracted the 2D kinematic map for the lensed quasar system \lensname\ from the JWST NIRSpec IFS, which is a first for a time-delay lens galaxy. The 2D resolved kinematics will improve the Hubble constant measurement by breaking the MSD and the MAD, after combining them with the measured time delays, lens models, and estimation of the line-of-sight convergence. This combination and Hubble constant measurement will be presented in a future study (Wang et al., in preparation). Resolved 1D kinematics from the same NIRSpec dataset presented here have been used in the recent measurement of $H_0$ by the TDCOSMO collaboration \citep{TDCOSMO25}.

The JWST NIRSpec dataset analyzed in this paper is the first of its kind, presenting novel challenges in robustly extracting stellar kinematics from this dataset. The unprecedentedly high $S/N$ of this dataset at the corresponding wavelengths necessitated meticulous modeling of all components, namely the lens galaxy, the background quasar, and its host galaxy, to robustly extract the stellar kinematics of the lens galaxy. Our methodology for this joint modeling has been packaged for future use into the publicly available \textsc{squirrel} pipeline, built upon the \textsc{pPXF} software program. Additionally, the JWST dataset required special processing beyond the standard JWST pipeline to mitigate several sources of non-random noise and artifacts, including uncleaned residual CRs and resampling noise, which manifests as "wiggles" in the spectra. Our JWST data reduction pipeline, \textsc{RegalJumper}, is also made publicly available. These methodological frameworks will enable rapid analysis of current and future JWST datasets of time-delay lenses to achieve less than 2\% precision in the Hubble constant measurement in the near future \citep{Birrer21b}.


\begin{acknowledgements}
We thank Elizabeth Buckley-Geer and Cheryl Pavlovski for useful discussions that improved this work.
This work is based on observations made with the NASA/ESA/CSA James Webb Space Telescope. The data were obtained from the Mikulski Archive for Space Telescopes at the Space Telescope Science Institute, which is operated by the Association of Universities for Research in Astronomy, Inc., under NASA contract NAS 5-03127 for JWST. These observations are associated with program \#1794. The specific observations analyzed can be accessed via \url{https://dx.doi.org/10.17909/wjva-0750}. Code scripts and kinematic maps are available from the corresponding author on request.
AJS received support from NASA through STScI grants JWST-GO-2974 and HST-GO-16773.
Support for this work was also provided by NASA through the NASA Hubble Fellowship grant HST-HF2-51492, awarded to AJS by the Space Telescope Science Institute, which is operated by the Association of Universities for Research in Astronomy, Inc., for NASA, under contract NAS5-26555.
TT acknowledges support by NSF through grants NSF-AST-1836016, NSF-AST-1906976, and NSF-AST-2407277, and from the Moore Foundation through grant 8548.
SHS thanks the Max Planck Society for support through the Max Planck Fellowship. SHS and HW are supported in part by the Deutsche Forschungsgemeinschaft (DFG, German Research Foundation) under Germany's Excellence Strategy - EXC-2094 - 390783311. 
This project has received funding from SNSF and the European Research Council (ERC) under the European Union’s Horizon 2020 research and innovation programme (COSMICLENS: grant agreement No.~787886; LENSNOVA: grant agreement No.~771776).
CDF acknowledges support for this work from the National Science Foundation under Grant No. AST-2407278.
\\
This research made use of \textsc{pPXF} \citep{Cappellari17, Cappellari23}, \textsc{pafit} \citep{Krajnovic06}, \textsc{vorbin} \citep{Cappellari03}, \textsc{lenstronomy} \citep{Birrer18, Birrer21b}, \textsc{stpsf} \citep{Perrin14}, \textsc{numpy} \citep{Oliphant15}, \textsc{scipy} \citep{Jones01}, \textsc{astropy} \citep{AstropyCollaboration13, AstropyCollaboration18}, \textsc{jupyter} \citep{Kluyver16}, \textsc{matplotlib} \citep{Hunter07}, \textsc{seaborn} \citep{Waskom14}, \textsc{emcee} \citep{Foreman-Mackey13}, and \textsc{getdist} (\url{https://github.com/cmbant/getdist}).
\end{acknowledgements}

%
\bibliographystyle{aa} 
\bibliography{ajshajib} 

\begin{thebibliography}{84}
\expandafter\ifx\csname natexlab\endcsname\relax\def\natexlab#1{#1}\fi

\bibitem[{Abdalla {et~al.}(2022)Abdalla, Abell{\'a}n, Aboubrahim, Agnello, Akarsu, Akrami, Alestas, Aloni, Amendola, Anchordoqui, Anderson, Arendse, Asgari, Ballardini, Barger, Basilakos, Batista, Battistelli, Battye, Benetti, Benisty, Berlin, {de Bernardis}, Berti, Bidenko, Birrer, Blakeslee, Boddy, Bom, Bonilla, Borghi, Bouchet, Braglia, Buchert, {Buckley-Geer}, Calabrese, Caldwell, Camarena, Capozziello, Casertano, Chen, Chluba, Chen, Chen, Chudaykin, Cicoli, Copi, Courbin, {Cyr-Racine}, Czerny, Dainotti, D'Amico, Davis, {de Cruz P{\'e}rez}, {de Haro}, Delabrouille, Denton, Dhawan, Dienes, Di~Valentino, Du, Eckert, {Escamilla-Rivera}, Fert{\'e}, Finelli, Fosalba, Freedman, Frusciante, Gazta{\~n}aga, Giar{\`e}, Giusarma, {G{\'o}mez-Valent}, Handley, Harrison, Hart, Hazra, Heavens, Heinesen, Hildebrandt, Hill, Hogg, Holz, Hooper, Hosseininejad, Huterer, Ishak, Ivanov, Jaffe, Jang, Jedamzik, Jimenez, Joseph, Joudaki, Kamionkowski, Karwal, Kazantzidis, Keeley, Klasen, Komatsu, Koopmans, Kumar, Lamagna, Lazkoz,
  Lee, Lesgourgues, Levi~Said, Lewis, L'Huillier, Lucca, Maartens, Macri, Marfatia, Marra, Martins, Masi, Matarrese, Mazumdar, Melchiorri, Mena, {Mersini-Houghton}, Mertens, Milakovi{\'c}, Minami, Miranda, {Moreno-Pulido}, Moresco, Mota, Mottola, Mozzon, Muir, Mukherjee, Mukherjee, Naselsky, Nath, Nesseris, Niedermann, Notari, Nunes, {\'O}~Colg{\'a}in, Owens, {\"O}z{\"u}lker, Pace, Paliathanasis, Palmese, Pan, Paoletti, Perez~Bergliaffa, Perivolaropoulos, Pesce, Pettorino, Philcox, Pogosian, Poulin, Poulot, Raveri, Reid, Renzi, Riess, Sabla, Salucci, Salzano, Saridakis, Sathyaprakash, Schmaltz, Sch{\"o}neberg, Scolnic, Sen, Sehgal, Shafieloo, {Sheikh-Jabbari}, Silk, Silvestri, Skara, Sloth, {Soares-Santos}, Sol{\`a}~Peracaula, Songsheng, Soriano, Staicova, Starkman, Szapudi, Teixeira, Thomas, Treu, Trott, {van de Bruck}, Vazquez, Verde, Visinelli, Wang, Wang, Wang, Watkins, Watson, Webb, Weiner, Weltman, Witte, Wojtak, Yadav, Yang, Zhao, \& Zumalac{\'a}rregui}]{Abdalla22}
Abdalla, E., Abell{\'a}n, G.~F., Aboubrahim, A., {et~al.} 2022, Journal of High Energy Astrophysics, 34, 49

\bibitem[{{Abdul Karim} {et~al.}(2025){Abdul Karim}, {Aguilar}, {Ahlen}, {Alam}, {Allen}, {Prieto}, {Alves}, {Anand}, {Andrade}, {Armengaud}, {Aviles}, {Bailey}, {Baltay}, {Bansal}, {Bault}, {Behera}, {BenZvi}, {Bianchi}, {Blake}, {Brieden}, {Brodzeller}, {Brooks}, {Buckley-Geer}, {Burtin}, {Calderon}, {Canning}, {Rosell}, {Carrilho}, {Casas}, {Castander}, {Charles}, {Chaussidon}, {Chaves-Montero}, {Chebat}, {Chen}, {Claybaugh}, {Cole}, {Cooper}, {Cuceu}, {Dawson}, {de la Macorra}, {de Mattia}, {Deiosso}, {Della Costa}, {Demina}, {Dey}, {Dey}, {Ding}, {Doel}, {Edelstein}, {Eisenstein}, {Elbers}, {Fagrelius}, {Fanning}, {Fern{\'a}ndez-Garc{\'\i}a}, {Ferraro}, {Font-Ribera}, {Forero-Romero}, {Frenk}, {Garcia-Quintero}, {Garrison}, {Gazta{\~n}aga}, {Gil-Mar{\'\i}n}, {Gontcho A Gontcho}, {Gonzalez}, {Gonzalez-Morales}, {Gordon}, {Green}, {Gutierrez}, {Guy}, {Hadzhiyska}, {Hahn}, {He}, {Herbold}, {Herrera-Alcantar}, {Ho}, {Honscheid}, {Howlett}, {Huterer}, {Ishak}, {Juneau}, {Kamble}, {Kara{\c{c}}ayl{\i}},
  {Kehoe}, {Kent}, {Kim}, {Kirkby}, {Kisner}, {Koposov}, {Kremin}, {Krolewski}, {Lahav}, {Lamman}, {Landriau}, {Lang}, {Lasker}, {Le Goff}, {Le Guillou}, {Leauthaud}, {Levi}, {Li}, {Li}, {Lodha}, {Lokken}, {Lozano-Rodr{\'\i}guez}, {Magneville}, {Manera}, {Martini}, {Matthewson}, {Meisner}, {Mena-Fern{\'a}ndez}, {Menegas}, {Mergulh{\~a}o}, {Miquel}, {Moustakas}, {Mu{\~n}oz-Guti{\'e}rrez}, {Mu{\~n}oz-Santos}, {Myers}, {Nadathur}, {Naidoo}, {Napolitano}, {Newman}, {Niz}, {Noriega}, {Paillas}, {Palanque-Delabrouille}, {Pan}, {Peacock}, {Pellejero Ibanez}, {Percival}, {P{\'e}rez-Fern{\'a}ndez}, {P{\'e}rez-R{\`a}fols}, {Pieri}, {Poppett}, {Prada}, {Rabinowitz}, {Raichoor}, {Ram{\'\i}rez-P{\'e}rez}, {Rashkovetskyi}, {Ravoux}, {Rich}, {Rocher}, {Rockosi}, {Rohlf}, {Rom{\'a}n-Herrera}, {Ross}, {Rossi}, {Ruggeri}, {Ruhlmann-Kleider}, {Samushia}, {Sanchez}, {Sanders}, {Schlegel}, {Schubnell}, {Seo}, {Shafieloo}, {Sharples}, {Silber}, {Sinigaglia}, {Sprayberry}, {Tan}, {Tarl{\'e}}, {Taylor}, {Turner},
  {Ure{\~n}a-L{\'o}pez}, {Vaisakh}, {Valdes}, {Valogiannis}, {Vargas-Maga{\~n}a}, {Verde}, {Walther}, {Weaver}, {Weinberg}, {White}, {Wolfson}, {Y{\`e}che}, {Yu}, {Zaborowski}, {Zarrouk}, {Zhai}, {Zhang}, {Zhao}, {Zhao}, {Zhou}, {Zou}, \& {DESI Collaboration}}]{DESI25}
{Abdul Karim}, M., {Aguilar}, J., {Ahlen}, S., {et~al.} 2025, \prd, 112, 083515

\bibitem[{{Astropy Collaboration}(2013)}]{AstropyCollaboration13}
{Astropy Collaboration}. 2013, \aap, 558, A33

\bibitem[{{Astropy Collaboration}(2018)}]{AstropyCollaboration18}
{Astropy Collaboration}. 2018, \aj, 156, 123

\bibitem[{{Astropy Collaboration} {et~al.}(2022){Astropy Collaboration}, {Price-Whelan}, Lim, Earl, Starkman, Bradley, Shupe, Patil, Corrales, Brasseur, N{\"o}the, Donath, Tollerud, Morris, Ginsburg, Vaher, Weaver, Tocknell, Jamieson, {van Kerkwijk}, Robitaille, Merry, Bachetti, G{\"u}nther, Aldcroft, {Alvarado-Montes}, Archibald, B{\'o}di, Bapat, Barentsen, Baz{\'a}n, Biswas, Boquien, Burke, Cara, Cara, Conroy, Conseil, Craig, Cross, Cruz, D'Eugenio, Dencheva, Devillepoix, Dietrich, Eigenbrot, Erben, Ferreira, {Foreman-Mackey}, Fox, Freij, Garg, Geda, Glattly, Gondhalekar, Gordon, Grant, Greenfield, Groener, Guest, Gurovich, Handberg, Hart, {Hatfield-Dodds}, Homeier, Hosseinzadeh, Jenness, Jones, Joseph, Kalmbach, Karamehmetoglu, Ka{\l}uszy{\'n}ski, Kelley, Kern, Kerzendorf, Koch, Kulumani, Lee, Ly, Ma, MacBride, Maljaars, Muna, Murphy, Norman, O'Steen, Oman, Pacifici, Pascual, {Pascual-Granado}, Patil, Perren, Pickering, Rastogi, Roulston, Ryan, Rykoff, Sabater, Sakurikar, Salgado, Sanghi, Saunders,
  Savchenko, Schwardt, {Seifert-Eckert}, Shih, Jain, Shukla, Sick, Simpson, Singanamalla, Singer, Singhal, Sinha, Sip{\H o}cz, Spitler, Stansby, Streicher, {\v S}umak, Swinbank, Taranu, Tewary, Tremblay, {de Val-Borro}, Van~Kooten, Vasovi{\'c}, Verma, {de Miranda Cardoso}, Williams, Wilson, Winkel, {Wood-Vasey}, Xue, Yoachim, Zhang, Zonca, \& {Astropy Project Contributors}}]{AstropyCollaboration22}
{Astropy Collaboration}, {Price-Whelan}, A.~M., Lim, P.~L., {et~al.} 2022, \apj, 935, 167

\bibitem[{Barth {et~al.}(2002)Barth, Ho, \& Sargent}]{Barth02}
Barth, A.~J., Ho, L.~C., \& Sargent, W. L.~W. 2002, The Astronomical Journal, 124, 2607

\bibitem[{Beifiori {et~al.}(2011)Beifiori, Maraston, Thomas, \& Johansson}]{Beifiori11}
Beifiori, A., Maraston, C., Thomas, D., \& Johansson, J. 2011, \aap, 531, A109

\bibitem[{Berge {et~al.}(2019)Berge, Massey, Baghi, \& Touboul}]{Berge19}
Berge, J., Massey, R., Baghi, Q., \& Touboul, P. 2019, \mnras, 486, 544

\bibitem[{Binney \& Mamon(1982)}]{Binney82}
Binney, J. \& Mamon, G.~A. 1982, \mnras, 200, 361

\bibitem[{Birrer(2021)}]{Birrer21}
Birrer, S. 2021, \apj, 919, 38

\bibitem[{Birrer \& Amara(2018)}]{Birrer18}
Birrer, S. \& Amara, A. 2018, Physics of the Dark Universe, 22, 189

\bibitem[{Birrer {et~al.}(2015)Birrer, Amara, \& Refregier}]{Birrer15}
Birrer, S., Amara, A., \& Refregier, A. 2015, \apj, 813, 102

\bibitem[{Birrer {et~al.}(2016)Birrer, Amara, \& Refregier}]{Birrer16}
Birrer, S., Amara, A., \& Refregier, A. 2016, \jcap, 8, 020

\bibitem[{Birrer {et~al.}(2024)Birrer, Millon, Sluse, Shajib, Courbin, Erickson, Koopmans, Suyu, \& Treu}]{Birrer24}
Birrer, S., Millon, M., Sluse, D., {et~al.} 2024, Space Science Reviews, 220, 48

\bibitem[{Birrer {et~al.}(2020)Birrer, Shajib, Galan, Millon, Treu, Agnello, Auger, Chen, Christensen, Collett, Courbin, Fassnacht, Koopmans, Marshall, Park, Rusu, Sluse, Spiniello, Suyu, {Wagner-Carena}, Wong, Barnab{\`e}, Bolton, Czoske, Ding, Frieman, \& de~Vyvere}]{Birrer20}
Birrer, S., Shajib, A.~J., Galan, A., {et~al.} 2020, \aap, 643, A165

\bibitem[{Birrer {et~al.}(2021)Birrer, Shajib, Gilman, Galan, Aalbers, Millon, Morgan, Pagano, Park, Teodori, Tessore, Ueland, de~Vyvere, {Wagner-Carena}, Wempe, Yang, Ding, Schmidt, Sluse, Zhang, \& Amara}]{Birrer21a}
Birrer, S., Shajib, A.~J., Gilman, D., {et~al.} 2021, JOSS, 6, 3283

\bibitem[{Birrer \& Treu(2021)}]{Birrer21b}
Birrer, S. \& Treu, T. 2021, \aap, 649, A61

\bibitem[{Bolton {et~al.}(2006)Bolton, Burles, Koopmans, Treu, \& Moustakas}]{Bolton06}
Bolton, A.~S., Burles, S., Koopmans, L. V.~E., Treu, T., \& Moustakas, L.~A. 2006, \apj, 638, 703

\bibitem[{Bolton {et~al.}(2012)Bolton, Schlegel, Aubourg, Bailey, Bhardwaj, Brownstein, Burles, Chen, Dawson, Eisenstein, Gunn, Knapp, Loomis, Lupton, Maraston, Muna, Myers, Olmstead, Padmanabhan, P{\^a}ris, Percival, Petitjean, Rockosi, Ross, Schneider, Shu, Strauss, Thomas, Tremonti, Wake, Weaver, \& {Wood-Vasey}}]{Bolton12}
Bolton, A.~S., Schlegel, D.~J., Aubourg, {\'E}., {et~al.} 2012, The Astronomical Journal, 144, 144

\bibitem[{Cappellari(2017)}]{Cappellari17}
Cappellari, M. 2017, \mnras, 466, 798

\bibitem[{Cappellari(2023)}]{Cappellari23}
Cappellari, M. 2023, \mnras, 526, 3273

\bibitem[{Cappellari \& Copin(2003)}]{Cappellari03}
Cappellari, M. \& Copin, Y. 2003, \mnras, 342, 345

\bibitem[{Cappellari {et~al.}(2013)Cappellari, Scott, Alatalo, Blitz, Bois, Bournaud, Bureau, Crocker, Davies, Davis, {de Zeeuw}, Duc, Emsellem, Khochfar, Krajnovi{\'c}, Kuntschner, McDermid, Morganti, Naab, Oosterloo, Sarzi, Serra, Weijmans, \& Young}]{Cappellari13}
Cappellari, M., Scott, N., Alatalo, K., {et~al.} 2013, \mnras, 432, 1709

\bibitem[{Cenarro {et~al.}(2001)Cenarro, Cardiel, Gorgas, Peletier, Vazdekis, \& Prada}]{Cenarro01}
Cenarro, A.~J., Cardiel, N., Gorgas, J., {et~al.} 2001, \mnras, 326, 959

\bibitem[{Chen {et~al.}(2016)Chen, Suyu, Wong, Fassnacht, Chiueh, Halkola, Hu, Auger, Koopmans, Lagattuta, McKean, \& Vegetti}]{Chen16}
Chen, G. C.-F., Suyu, S.~H., Wong, K.~C., {et~al.} 2016, \mnras, 462, 3457

\bibitem[{Claeskens {et~al.}(2006)Claeskens, Sluse, Riaud, \& Surdej}]{Claeskens06}
Claeskens, J.-F., Sluse, D., Riaud, P., \& Surdej, J. 2006, \aap, 451, 865

\bibitem[{D'Eugenio {et~al.}(2024)D'Eugenio, {P{\'e}rez-Gonz{\'a}lez}, Maiolino, Scholtz, Perna, Circosta, {\"U}bler, Arribas, B{\"o}ker, Bunker, Carniani, Charlot, Chevallard, Cresci, {Curtis-Lake}, Jones, Kumari, Lamperti, Looser, Parlanti, Rix, Robertson, Rodr{\'i}guez Del~Pino, Tacchella, Venturi, \& Willott}]{DEugenio24}
D'Eugenio, F., {P{\'e}rez-Gonz{\'a}lez}, P.~G., Maiolino, R., {et~al.} 2024, Nature Astronomy, 8, 1443

\bibitem[{{Di Valentino} {et~al.}(2025){Di Valentino}, {Said}, {Riess}, {Pollo}, {Poulin}, {G{\'o}mez-Valent}, {Weltman}, {Palmese}, {Huang}, {van de Bruck}, {Saraf}, {Kuo}, {Uhlemann}, {Grand{\'o}n}, {Paz}, {Eckert}, {Teixeira}, {Saridakis}, {Colg{\'a}in}, {Beutler}, {Niedermann}, {Bajardi}, {Barenboim}, {Gubitosi}, {Musella}, {Banik}, {Szapudi}, {Singal}, {Cases}, {Chluba}, {Torrado}, {Mifsud}, {Jedamzik}, {Said}, {Dialektopoulos}, {Herold}, {Perivolaropoulos}, {Zu}, {Galbany}, {Breuval}, {Visinelli}, {Escamilla}, {Anchordoqui}, {Sheikh-Jabbari}, {Lembo}, {Dainotti}, {Vincenzi}, {Asgari}, {Gerbino}, {Forconi}, {Cantiello}, {Moresco}, {Benetti}, {Sch{\"o}neberg}, {Akarsu}, {Nunes}, {Bernardo}, {Ch{\'a}vez}, {Anderson}, {Watkins}, {Capozziello}, {Li}, {Vagnozzi}, {Pan}, {Treu}, {Irsic}, {Handley}, {Giar{\`e}}, {Murakami}, {Banihashemi}, {Poudou}, {Heavens}, {Kogut}, {Domi}, {Lenart}, {Melchiorri}, {Vadal{\`a}}, {Amon}, {Rivera}, {Reeves}, {Zhuk}, {Bonanno}, {{\"O}vg{\"u}n}, {Pisani}, {Talebian}, {Abebe},
  {Aboubrahim}, {Gonz{\'a}lez Mor{\'a}n}, {Kov{\'a}cs}, {Lymperis}, {Papatriantafyllou}, {Liddle}, {Paliathanasis}, {Borowiec}, {Yadav}, {Yadav}, {Sen}, {William}, {Davis}, {Shajib}, {Walters}, {Lonappan}, {Chudaykin}, {Capodagli}, {da Silva}, {De Felice}, {Racioppi}, {Oficial}, {Montiel}, {Favale}, {Bernui}, {Velasco}, {Heinesen}, {Bakopoulos}, {Chatzistavrakidis}, {Khanpour}, {Sathyaprakash}, {Zgirski}, {L'Huillier}, {Famaey}, {Jain}, {Zhang}, {Karmakar}, {Dragovich}, {Thomas}, {Correa}, {Boiza}, {Marques}, {Escamilla-Rivera}, {Tzerefos}, {Zhang}, {De Leo}, {Pfeifer}, {Lee}, {Venter}, {Gomes}, {Roque De bom}, {Moreno-Pulido}, {Iosifidis}, {Grin}, {Blixt}, {Scolnic}, {Oriti}, {Dobrycheva}, {Bettoni}, {Benisty}, {Fern{\'a}ndez-Arenas}, {Wiltshire}, {Sanchez Cid}, {Tamayo}, {Valls-Gabaud}, {Pedrotti}, {Wang}, {Staicova}, {Totolou}, {Rubiera-Garcia}, {Milakovi{\'c}}, {Pesce}, {Sluse}, {Borka}, {Yusofi}, {Giusarma}, {Terlevich}, {Tomasetti}, {Vagenas}, {Fazzari}, {Ferreira}, {Barakovic}, {Dimastrogiovanni},
  {Holm}, {Mottola}, {{\"O}z{\"u}lker}, {Specogna}, {Brocato}, {Jensko}, {Enriquez}, {Bhatia}, {Bresolin}, {Avila}, {Bouch{\`e}}, {Bombacigno}, {Anagnostopoulos}, {Pace}, {Sorrenti}, {Lobo}, {Courbin}, {Hansen}, {Sloan}, {Farrugia}, {Lynch}, {Garcia-Arroyo}, {Raimondo}, {Lambiase}, {Anand}, {Poulot}, {Leon}, {Kouniatalis}, {Nardini}, {Cs{\"o}rnyei}, \& {Galloni}}]{DiValentino25}
{Di Valentino}, E., {Said}, J.~L., {Riess}, A., {et~al.} 2025, Physics of the Dark Universe, 49, 101965

\bibitem[{{Dumont} {et~al.}(2025){Dumont}, {Neumayer}, {Seth}, {B{\"o}ker}, {Eracleous}, {Goold}, {Greene}, {G{\"u}ltekin}, {Ho}, {Walsh}, \& {L{\"u}tzgendorf}}]{Dumont25}
{Dumont}, A., {Neumayer}, N., {Seth}, A.~C., {et~al.} 2025, \aap, 703, A54

\bibitem[{Eigenbrod {et~al.}(2005)Eigenbrod, Courbin, Vuissoz, Meylan, Saha, \& Dye}]{Eigenbrod05}
Eigenbrod, A., Courbin, F., Vuissoz, C., {et~al.} 2005, \aap, 436, 25

\bibitem[{Falco {et~al.}(1985)Falco, Gorenstein, \& Shapiro}]{Falco85}
Falco, E.~E., Gorenstein, M.~V., \& Shapiro, I.~I. 1985, \apjl, 289, L1

\bibitem[{{Foreman-Mackey} {et~al.}(2013){Foreman-Mackey}, Hogg, Lang, \& Goodman}]{Foreman-Mackey13}
{Foreman-Mackey}, D., Hogg, D.~W., Lang, D., \& Goodman, J. 2013, \pasp, 125, 306

\bibitem[{Gavazzi {et~al.}(2012)Gavazzi, Treu, Marshall, Brault, \& Ruff}]{Gavazzi12}
Gavazzi, R., Treu, T., Marshall, P.~J., Brault, F., \& Ruff, A. 2012, \apj, 761, 170

\bibitem[{Guo {et~al.}(2018)Guo, Shen, \& Wang}]{Guo18}
Guo, H., Shen, Y., \& Wang, S. 2018, Astrophysics Source Code Library, ascl:1809.008

\bibitem[{Humphrey \& Buote(2010)}]{Humphrey10}
Humphrey, P.~J. \& Buote, D.~A. 2010, \mnras, 403, 2143

\bibitem[{Hunter(2007)}]{Hunter07}
Hunter, J.~D. 2007, Computing in Science and Engineering, 9, 90

\bibitem[{Hutsem{\'e}kers {et~al.}(2024)Hutsem{\'e}kers, Sluse, \& Savi{\'c}}]{Hutsemekers24}
Hutsem{\'e}kers, D., Sluse, D., \& Savi{\'c}, {\DJ}. 2024, \aap, 691, A292

\bibitem[{Jones {et~al.}(2001)Jones, Oliphant, Peterson, \& {Others}}]{Jones01}
Jones, E., Oliphant, T., Peterson, P., \& {Others}. 2001, {{SciPy}}: {{Open}} Source Scientific Tools for {{Python}}

\bibitem[{Kluyver {et~al.}(2016)Kluyver, {Ragan-Kelley}, P{\'e}rez, Granger, Bussonnier, Frederic, Kelley, Hamrick, Grout, Corlay, Ivanov, Avila, Abdalla, \& Willing}]{Kluyver16}
Kluyver, T., {Ragan-Kelley}, B., P{\'e}rez, F., {et~al.} 2016, in Positioning and {{Power}} in {{Academic Publishing}}: {{Players}}, {{Agents}} and {{Agendas}}, ed. F.~Loizides \& B.~Schmidt (IOS Press BV, Amsterdam, Netherlands), 87--90

\bibitem[{{Knabel} {et~al.}(2025{\natexlab{a}}){Knabel}, {Mozumdar}, {Shajib}, {Treu}, {Cappellari}, {Spiniello}, \& {Birrer}}]{Knabel25}
{Knabel}, S., {Mozumdar}, P., {Shajib}, A.~J., {et~al.} 2025{\natexlab{a}}, \aap, 703, A117

\bibitem[{{Knabel} {et~al.}(2025{\natexlab{b}}){Knabel}, {Treu}, {Cappellari}, {Shajib}, {Chen}, {Birrer}, \& {Bennert}}]{Knabel24}
{Knabel}, S., {Treu}, T., {Cappellari}, M., {et~al.} 2025{\natexlab{b}}, \apj, 990, 51

\bibitem[{Knox \& Millea(2020)}]{Knox19}
Knox, L. \& Millea, M. 2020, \prd, 101, 043533

\bibitem[{Krajnovi{\'c} {et~al.}(2006)Krajnovi{\'c}, Cappellari, {de Zeeuw}, \& Copin}]{Krajnovic06}
Krajnovi{\'c}, D., Cappellari, M., {de Zeeuw}, P.~T., \& Copin, Y. 2006, \mnras, 366, 787

\bibitem[{Lagattuta {et~al.}(2012)Lagattuta, Vegetti, Fassnacht, Auger, Koopmans, \& McKean}]{Lagattuta12}
Lagattuta, D.~J., Vegetti, S., Fassnacht, C.~D., {et~al.} 2012, \mnras, 424, 2800

\bibitem[{Law {et~al.}(2023)Law, E.~Morrison, Argyriou, Patapis, {\'A}lvarez-M{\'a}rquez, Labiano, \& Vandenbussche}]{Law23}
Law, D.~R., E.~Morrison, J., Argyriou, I., {et~al.} 2023, The Astronomical Journal, 166, 45

\bibitem[{Linder(2011)}]{Linder11}
Linder, E.~V. 2011, \prd, 84, 123529

\bibitem[{{Lynch} {et~al.}(2024){Lynch}, {Knox}, \& {Chluba}}]{Lynch24}
{Lynch}, G.~P., {Knox}, L., \& {Chluba}, J. 2024, \prd, 110, 083538

\bibitem[{Millon {et~al.}(2020)Millon, Galan, Courbin, Treu, Suyu, Ding, Birrer, Chen, Shajib, Sluse, Wong, Agnello, Auger, {Buckley-Geer}, Chan, Collett, Fassnacht, Hilbert, Koopmans, Motta, Mukherjee, Rusu, Sonnenfeld, Spiniello, \& {Van de Vyvere}}]{Millon20a}
Millon, M., Galan, A., Courbin, F., {et~al.} 2020, \aap, 639, A101

\bibitem[{Navarro {et~al.}(1997)Navarro, Frenk, \& White}]{Navarro97}
Navarro, J.~F., Frenk, C.~S., \& White, S. D.~M. 1997, \apj, 490, 493

\bibitem[{Oliphant(2015)}]{Oliphant15}
Oliphant, T.~E. 2015, Guide to {{NumPy}}, 2nd edn. (USA: CreateSpace Independent Publishing Platform)

\bibitem[{Perna {et~al.}(2023)Perna, Arribas, Marshall, D'Eugenio, {\"U}bler, Bunker, Charlot, Carniani, Jakobsen, Maiolino, Rodr{\'i}guez Del~Pino, Willott, B{\"o}ker, Circosta, Cresci, Curti, Husemann, Kumari, Lamperti, {P{\'e}rez-Gonz{\'a}lez}, \& Scholtz}]{Perna23}
Perna, M., Arribas, S., Marshall, M., {et~al.} 2023, \aap, 679, A89

\bibitem[{Perrin {et~al.}(2014)Perrin, Sivaramakrishnan, Lajoie, Elliott, Pueyo, Ravindranath, \& Albert}]{Perrin14}
Perrin, M.~D., Sivaramakrishnan, A., Lajoie, C.-P., {et~al.} 2014, in Space {{Telescopes}} and {{Instrumentation}} 2014: {{Optical}}, {{Infrared}}, and {{Millimeter Wave}}, Vol. 9143, 91433X

\bibitem[{Rauscher(2024)}]{Rauscher24}
Rauscher, B.~J. 2024, Publications of the Astronomical Society of the Pacific, 136, 015001

\bibitem[{Refregier(2003)}]{Refregier03}
Refregier, A. 2003, \mnras, 338, 35

\bibitem[{Riess {et~al.}(2022)Riess, Yuan, Macri, Scolnic, Brout, Casertano, Jones, Murakami, Anand, Breuval, Brink, Filippenko, Hoffmann, Jha, D'arcy~Kenworthy, Mackenty, Stahl, \& Zheng}]{Riess22}
Riess, A.~G., Yuan, W., Macri, L.~M., {et~al.} 2022, \apj, 934, L7

\bibitem[{S{\'e}rsic(1968)}]{Sersic68}
S{\'e}rsic, J.~L. 1968, Atlas de {{Galaxias Australes}}

\bibitem[{{Shajib}(2025)}]{Shajib25b}
{Shajib}, A.~J. 2025, Under review by JOSS, arXiv:2507.13341

\bibitem[{Shajib {et~al.}(2019)Shajib, Birrer, Treu, Auger, Agnello, Anguita, {Buckley-Geer}, Chan, Collett, Courbin, Fassnacht, Frieman, Kayo, Lemon, Lin, Marshall, McMahon, More, Morgan, Motta, Oguri, Ostrovski, Rusu, Schechter, Shanks, Suyu, Meylan, Abbott, Allam, Annis, Avila, Bertin, Brooks, Carnero~Rosell, Carrasco~Kind, Carretero, Cunha, {da Costa}, De~Vicente, Desai, Doel, Flaugher, Fosalba, {Garc{\'i}a-Bellido}, Gerdes, Gruen, Gruendl, Gutierrez, Hartley, Hollowood, Hoyle, James, Kuehn, Kuropatkin, Lahav, Lima, Maia, March, Marshall, Melchior, Menanteau, Miquel, Plazas, Sanchez, Scarpine, {Sevilla-Noarbe}, Smith, {Soares-Santos}, Sobreira, Suchyta, Swanson, Tarle, \& Walker}]{Shajib19}
Shajib, A.~J., Birrer, S., Treu, T., {et~al.} 2019, \mnras, 483, 5649

\bibitem[{Shajib {et~al.}(2023)Shajib, Mozumdar, Chen, Treu, Cappellari, Knabel, Suyu, Bennert, Frieman, Sluse, Birrer, Courbin, Fassnacht, Villafa{\~n}a, \& Williams}]{Shajib23}
Shajib, A.~J., Mozumdar, P., Chen, G. C.~F., {et~al.} 2023, \aap, 673, A9

\bibitem[{Shajib {et~al.}(2018)Shajib, Treu, \& Agnello}]{Shajib18}
Shajib, A.~J., Treu, T., \& Agnello, A. 2018, \mnras, 473, 210

\bibitem[{{Shajib} {et~al.}(2025){Shajib}, {Treu}, {Melo}, {Roberts-Borsani}, {Knabel}, {Cappellari}, \& {Frieman}}]{Shajib25a}
{Shajib}, A.~J., {Treu}, T., {Melo}, A., {et~al.} 2025, \aap, 702, L12

\bibitem[{Sluse {et~al.}(2007)Sluse, Claeskens, Hutsem{\'e}kers, \& Surdej}]{Sluse07}
Sluse, D., Claeskens, J.-F., Hutsem{\'e}kers, D., \& Surdej, J. 2007, \aap, 468, 885

\bibitem[{Sluse {et~al.}(2003)Sluse, Surdej, Claeskens, Hutsem{\'e}kers, Jean, Courbin, Nakos, Billeres, \& Khmil}]{Sluse03}
Sluse, D., Surdej, J., Claeskens, J.-F., {et~al.} 2003, \aap, 406, L43

\bibitem[{Suyu {et~al.}(2013)Suyu, Auger, Hilbert, Marshall, Tewes, Treu, Fassnacht, Koopmans, Sluse, Blandford, Courbin, \& Meylan}]{Suyu13}
Suyu, S.~H., Auger, M.~W., Hilbert, S., {et~al.} 2013, \apj, 766, 70

\bibitem[{Suyu {et~al.}(2017)Suyu, Bonvin, Courbin, Fassnacht, Rusu, Sluse, Treu, Wong, Auger, Ding, Hilbert, Marshall, Rumbaugh, Sonnenfeld, Tewes, Tihhonova, Agnello, Blandford, Chen, Collett, Koopmans, Liao, Meylan, \& Spiniello}]{Suyu17}
Suyu, S.~H., Bonvin, V., Courbin, F., {et~al.} 2017, \mnras, 468, 2590

\bibitem[{Suyu {et~al.}(2014)Suyu, Treu, Hilbert, Sonnenfeld, Auger, Blandford, Collett, Courbin, Fassnacht, Koopmans, Marshall, Meylan, Spiniello, \& Tewes}]{Suyu14}
Suyu, S.~H., Treu, T., Hilbert, S., {et~al.} 2014, \apjl, 788, L35

\bibitem[{{TDCOSMO Collaboration}(2025)}]{TDCOSMO25}
{TDCOSMO Collaboration}. 2025, \aap, 704, A63

\bibitem[{Tessore \& Metcalf(2015)}]{Tessore15}
Tessore, N. \& Metcalf, R.~B. 2015, \aap, 580, A79

\bibitem[{Tewes {et~al.}(2013)Tewes, Courbin, Meylan, Kochanek, Eulaers, Cantale, Mosquera, Magain, Van~Winckel, Sluse, Cataldi, V{\"o}r{\"o}s, \& Dye}]{Tewes13}
Tewes, M., Courbin, F., Meylan, G., {et~al.} 2013, \aap, 556, A22

\bibitem[{Treu {et~al.}(2018)Treu, Agnello, Baumer, Birrer, {Buckley-Geer}, Courbin, Kim, Lin, Marshall, Nord, Schechter, Sivakumar, Abramson, Anguita, Apostolovski, Auger, Chan, Chen, Collett, Fassnacht, Hsueh, Lemon, McMahon, Motta, Ostrovski, Rojas, Rusu, Williams, Frieman, Meylan, Suyu, Abbott, Abdalla, Allam, Annis, Avila, Banerji, Brooks, Carnero~Rosell, Carrasco~Kind, Carretero, Castander, D'Andrea, {da Costa}, De~Vicente, Doel, Eifler, Flaugher, Fosalba, {Garc{\'i}a-Bellido}, Goldstein, Gruen, Gruendl, Gutierrez, Hartley, Hollowood, Honscheid, James, Kuehn, Kuropatkin, Lima, Maia, Martini, Menanteau, Miquel, Plazas, Romer, Sanchez, Scarpine, Schindler, Schubnell, {Sevilla-Noarbe}, Smith, Smith, {Soares-Santos}, Sobreira, Suchyta, Swanson, Tarle, Thomas, Tucker, \& Walker}]{Treu18}
Treu, T., Agnello, A., Baumer, M.~A., {et~al.} 2018, \mnras, 481, 1041

\bibitem[{Treu {et~al.}(2016)Treu, Brammer, Diego, Grillo, Kelly, Oguri, Rodney, Rosati, Sharon, Zitrin, Balestra, Brada{\v c}, Broadhurst, Caminha, Halkola, Hoag, Ishigaki, Johnson, Karman, Kawamata, Mercurio, Schmidt, Strolger, Suyu, Filippenko, Foley, Jha, \& Patel}]{Treu16}
Treu, T., Brammer, G., Diego, J.~M., {et~al.} 2016, \apj, 817, 60

\bibitem[{Treu \& Koopmans(2002)}]{Treu02a}
Treu, T. \& Koopmans, L. V.~E. 2002, \mnras, 337, L6

\bibitem[{{Treu} \& {Shajib}(2024)}]{Treu23}
{Treu}, T. \& {Shajib}, A.~J. 2024, in The Hubble Constant Tension, ed. E.~{Di Valentino} \& {Brout Dillon}, 251--276

\bibitem[{Treu {et~al.}(2022)Treu, Suyu, \& Marshall}]{Treu22}
Treu, T., Suyu, S.~H., \& Marshall, P.~J. 2022, Strong Lensing Time-Delay Cosmography in the 2020s

\bibitem[{{Turner} {et~al.}(2024){Turner}, {Smith}, \& {Collett}}]{Turner24}
{Turner}, H.~C., {Smith}, R.~J., \& {Collett}, T.~E. 2024, \mnras, 528, 3559

\bibitem[{Vagnozzi(2023)}]{Vagnozzi23}
Vagnozzi, S. 2023, Universe, 9, 393

\bibitem[{Valdes {et~al.}(2004)Valdes, Gupta, Rose, Singh, \& Bell}]{Valdes04}
Valdes, F., Gupta, R., Rose, J.~A., Singh, H.~P., \& Bell, D.~J. 2004, The Astrophysical Journal Supplement Series, 152, 251

\bibitem[{{van Dokkum}(2001)}]{vanDokkum01}
{van Dokkum}, P.~G. 2001, Publications of the Astronomical Society of the Pacific, 113, 1420

\bibitem[{{Verro} {et~al.}(2022){Verro}, {Trager}, {Peletier}, {Lan{\c{c}}on}, {Gonneau}, {Vazdekis}, {Prugniel}, {Chen}, {Coelho}, {S{\'a}nchez-Bl{\'a}zquez}, {Martins}, {Arentsen}, {Lyubenova}, {Falc{\'o}n-Barroso}, \& {Dries}}]{Verro22}
{Verro}, K., {Trager}, S.~C., {Peletier}, R.~F., {et~al.} 2022, \aap, 660, A34

\bibitem[{{Wang} {et~al.}(2025){Wang}, {Suyu}, {Galan}, {Halkola}, {Cappellari}, {Shajib}, \& {Cernetic}}]{Wang25}
{Wang}, H., {Suyu}, S.~H., {Galan}, A., {et~al.} 2025, \aap, 701, A280

\bibitem[{Waskom {et~al.}(2014)Waskom, Botvinnik, Hobson, Cole, Halchenko, Hoyer, Miles, Augspurger, Yarkoni, Megies, Coelho, Wehner, {cynddl}, Ziegler, {diego0020}, Zaytsev, Hoppe, Seabold, Cloud, Koskinen, Meyer, Qalieh, \& Allan}]{Waskom14}
Waskom, M., Botvinnik, O., Hobson, P., {et~al.} 2014, Seaborn: V0.5.0 ({{November}} 2014)

\bibitem[{Westfall {et~al.}(2019)Westfall, Cappellari, Bershady, Bundy, Belfiore, Ji, Law, Schaefer, Shetty, Tremonti, Yan, Andrews, Brownstein, Cherinka, Coccato, Drory, Maraston, Parikh, {S{\'a}nchez-Gallego}, Thomas, Weijmans, {Barrera-Ballesteros}, Du, Goddard, Li, Masters, Ibarra~Medel, S{\'a}nchez, Yang, Zheng, \& Zhou}]{Westfall19}
Westfall, K.~B., Cappellari, M., Bershady, M.~A., {et~al.} 2019, The Astronomical Journal, 158, 231

\bibitem[{Y{\i}ld{\i}r{\i}m {et~al.}(2023)Y{\i}ld{\i}r{\i}m, Suyu, Chen, \& Komatsu}]{Yildirim23}
Y{\i}ld{\i}r{\i}m, A., Suyu, S.~H., Chen, G. C.~F., \& Komatsu, E. 2023, \aap, 675, A21

\bibitem[{Y{\i}ld{\i}r{\i}m {et~al.}(2020)Y{\i}ld{\i}r{\i}m, Suyu, \& Halkola}]{Yildirim20}
Y{\i}ld{\i}r{\i}m, A., Suyu, S.~H., \& Halkola, A. 2020, \mnras, 493, 4783

\end{thebibliography}
%

\appendix

\section{Correction for wiggles in spectra} \label{app:wiggle_correction}

\begin{figure*}[ht!]
    \centering
	\includegraphics[width=0.93\textwidth]{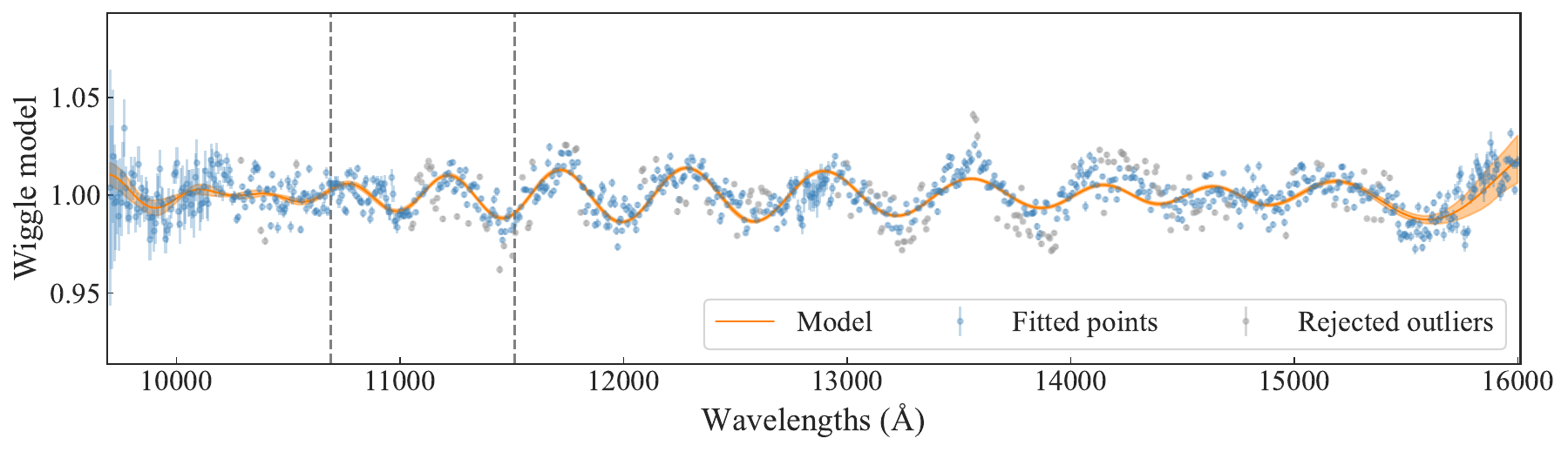}
	\caption{\label{fig:wiggle_model}
	Fit wiggle model used to correct for the resampling noise. The data points show the signal of wiggle $d(\lambda)/T(\lambda)$ for the central spaxel on the lens galaxy, where $T(\lambda)$ is obtained from the best-fit model $m(\lambda)$. The orange line shows the $W(\lambda)$ function in the best-fit $m(\lambda)$, with the orange shaded region illustrating the 1$\sigma$ confidence region. The vertical gray dashed lines show the wavelength range containing the \caii\ triplets of the lens galaxy that we fit to measure the lens galaxy's kinematics. The gray data points are rejected outliers using the false discovery rate method.}
\end{figure*}

In this appendix, we briefly describe the "wiggle" correction we performed on the reduced datacube spectra and provide the relevant settings. We used the software package \textsc{raccoon}, the detailed mechanism and features of which are presented by \shajibraccoont.

The wiggles or modulation on the spectra are artifacts due to resampling noise, as the pixels are not optimally sampled by the PSF, and our four-point dither pattern is insufficient to recover the optimal sampling \citep{Law23}. In \textsc{raccoon}, the wiggle is modeled as a sinusoidal chirp function:
\begin{equation}
	W(\lambda) = 1 + A(\lambda) \left[ \sin (\phi_\lambda) + k_1 \sin^2 (\phi_\lambda) + k_2 \sin(3\phi_\lambda) \right],
\end{equation}
where $A(\lambda)$ is the wavelength-dependent amplitude and $\phi_\lambda = \lambda f(\lambda) + \phi_0$ is wavelength-depedent phase term. The wiggle can be made asymmetric with the $k_1 \neq 0$ and the peaks can be symmetrically sharpened (or, de-sharpened) with $k_2 \neq 0$. In our use case, we set both $k_1 = 0$ and $k_2 = 0$.

The wiggles affect the single-spaxel spectra. However, they wash out if the spectra are summed within an aperture with a few spaxel radius. From our tests, we find that the wiggle signal is almost completely washed out with an aperture radius larger than three pixels, which approximately equals the FWHM of the PSF. We obtain the best-fit wiggle model by fitting the single-spaxel spectra $d(\lambda)$ with a model
\begin{equation}
	m (\lambda) = W(\lambda) \, T(\lambda),
\end{equation}
with the template $T(\lambda)$ for the original spectra constructed as
\begin{equation}
    T(\lambda) = c_1\,a(\lambda) + c_2 \, s(\lambda) + c_3 \, \lambda^b + c_4 \, P(\lambda),
\end{equation}
where $c_1,...,c_4$ are linear coefficients, $a(\lambda)$ is the aperture-summed spectra around the spaxel, $s(\lambda)$ is the spectra summed within a shell centered on the spaxel, and $P(\lambda)$ is a polynomial of chosen degree. The power-law term $\lambda^b$ and the polynomial $P(\lambda)$ are expected to model the change in the continuum level between the single-spaxel spectra and the aperture-summed one. The combination $c_1\,a(\lambda) + c_2 \, s(\lambda)$ allows accounting for changes in the emission or absorption line shapes between the modeled single-spaxel spectra and the aperture-summed one \citep{Dumont25}. In our case, we chose four spaxels as the radius of the aperture to extract $a(\lambda)$, the inner and outer radii of the shell to extract $s(\lambda)$ to be three and five spaxels, respectively, and the polynomial degree to be two. The functions $A(\lambda)$ and $\phi_\lambda$ were modeled with B-splines with three and seven knots (thus with five and nine free parameters), respectively. We propagate the uncertainty of the above fitting procedure of the wiggle model and add it to the noise levels of the corrected spectra $d(\lambda)/W(\lambda)$.

\begin{figure}
\includegraphics[width=\columnwidth]{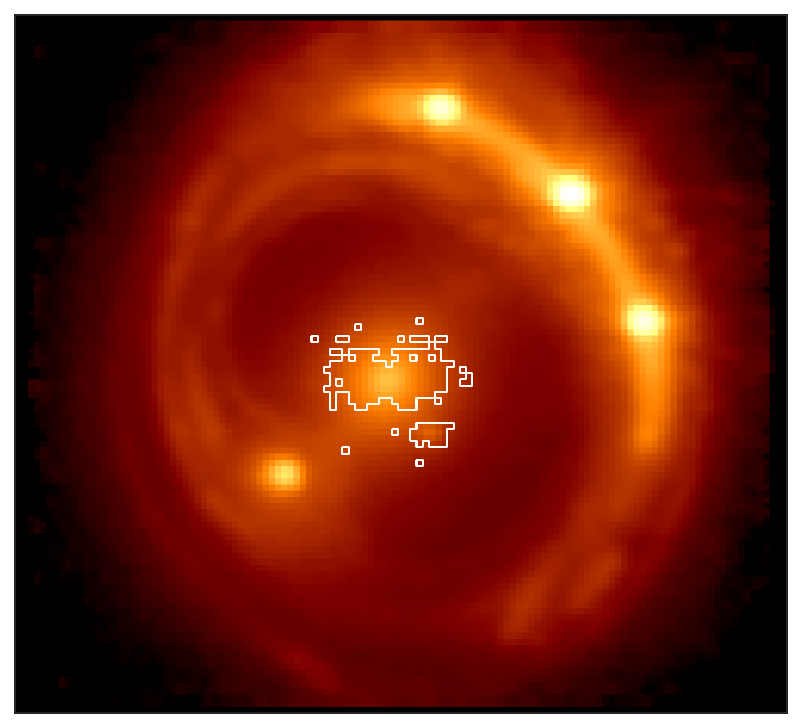}
\caption{\label{fig:wcleaned_pixels}
Spaxels cleaned for wiggles (within the regions enclosed with white borders). Only the spaxels within a circle of 0\farcs7 radius from the lens galaxy center were considered to be cleaned. Then, we only cleaned the spaxels for which the wiggle was detected at a confidence level of more than $3\sigma$ and the best-fit wiggle model explains more than half of the variance or scatter observed in the wiggle signal. The wiggles are more prominent in spaxels with more flux. As a result, the white contours above primarily encompass the central region of the lens galaxy and the satellite galaxy.
}
\end{figure}

\begin{figure}
	\includegraphics[width=\columnwidth]{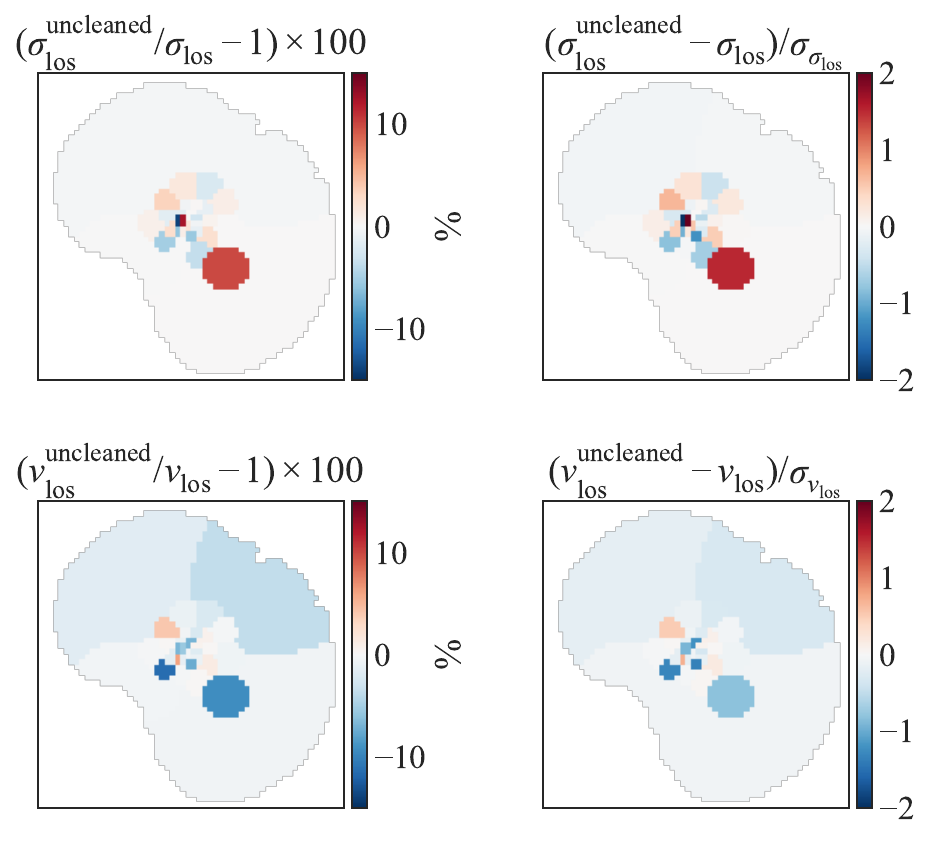}
	\caption{\label{fig:wclean_difference}
	Impact of wiggle cleaning on the measured kinematics. The illustrated test is for a single model setup out of the 48 combinations described in \secref{sec:systematic}. The first column displays the fractional differences in the measured dispersions (top row) and mean velocities (bottom row) within the Voronoi bins, while the second column shows these differences normalized by the statistical uncertainty.
	}
\end{figure}

We perform the cleaning for the spaxels within an aperture of 0\farcs7 radius centered on the lens galaxy, and for which the wiggle is detected with more than $3\sigma$ and the wiggle model explains more than half of the scatter observed in the "wiggle signal" [i.e., $d(\lambda)/T(\lambda)$]. We chose the 0\farcs7 radius, as this would be an \textit{a priori} safe choice for a radius outside of which all the spaxels would belong to Voronoi bins large enough to smooth out the wiggles after summing within. Although we fit within a narrow wavelength range out of the whole observed region, we fit and correct for the wiggles across the whole wavelength range so that the modeled wiggle signal is less affected by local outliers, noise, or scatter within the fit wavelength region, and the fit wiggle pattern is dictated more by a global pattern. In \figref{fig:wiggle_model}, we illustrate an example of the wiggle signal and the best-fit wiggle model. In \figref{fig:wcleaned_pixels}, we show the spaxels for which wiggle cleaning was performed. In \figref{fig:wclean_difference}, we illustrate the impact of the cleaning on the measured kinematics (here, only the systematics from the template library choice were marginalized). Without cleaning, the measured velocity dispersions would shift down by $\sim$0.6\% on average in the central region of the lens galaxy (that excludes the four outermost bins and the one containing the satellite), and the dispersion in the bin with the satellite galaxy would shift up by $\sim$10\%.

\section{Quasar contamination levels } \label{app:quasar_contamination}

\begin{figure}[!h]
	\includegraphics[width=\columnwidth]{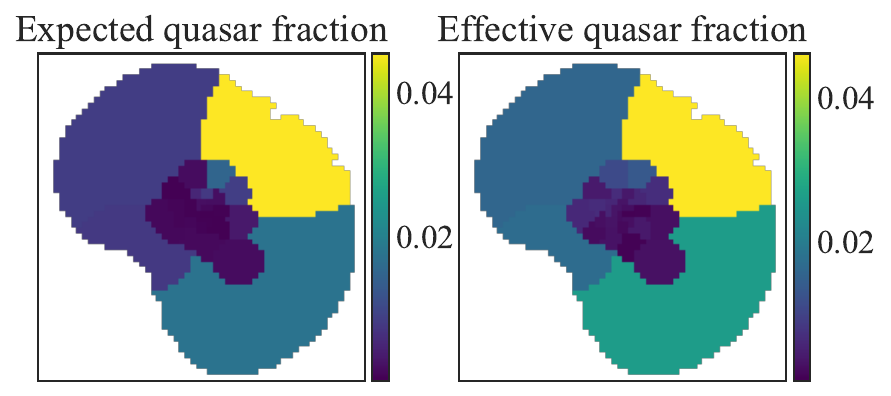}
	\caption{\label{fig:quasar_contamination}
	Expected quasar contamination in each Voronoi‑binned spectrum from the best‑fit lens model (left) and the effective levels from our \textsc{pPXF} fits (right). The effective contamination is consistent across bins and follows the same radial increase -- from low at the center to higher outward -- highlighting the effectiveness of our multi‑component fitting approach.
	}
\end{figure}

{In this appendix, we show the expected quasar contamination in each Voronoi‑binned spectrum alongside the effective levels recovered from our \textsc{pPXF} fits (\figref{fig:quasar_contamination}). The expected levels are derived from the decomposition based on our best‑fit lens model in \secref{sec:lens_modeling}. The effective contamination measured by \textsc{pPXF} closely matches these expectations and follows the same spatial trend -- minimal in the center and increasing toward the quasar image -- demonstrating the effectiveness of our multi‑component fitting approach.}

\section{Fits of individual Voronoi bins} \label{app:individual_voronoi_bin_fits}

\begin{figure*}
	\includegraphics[width=\textwidth]{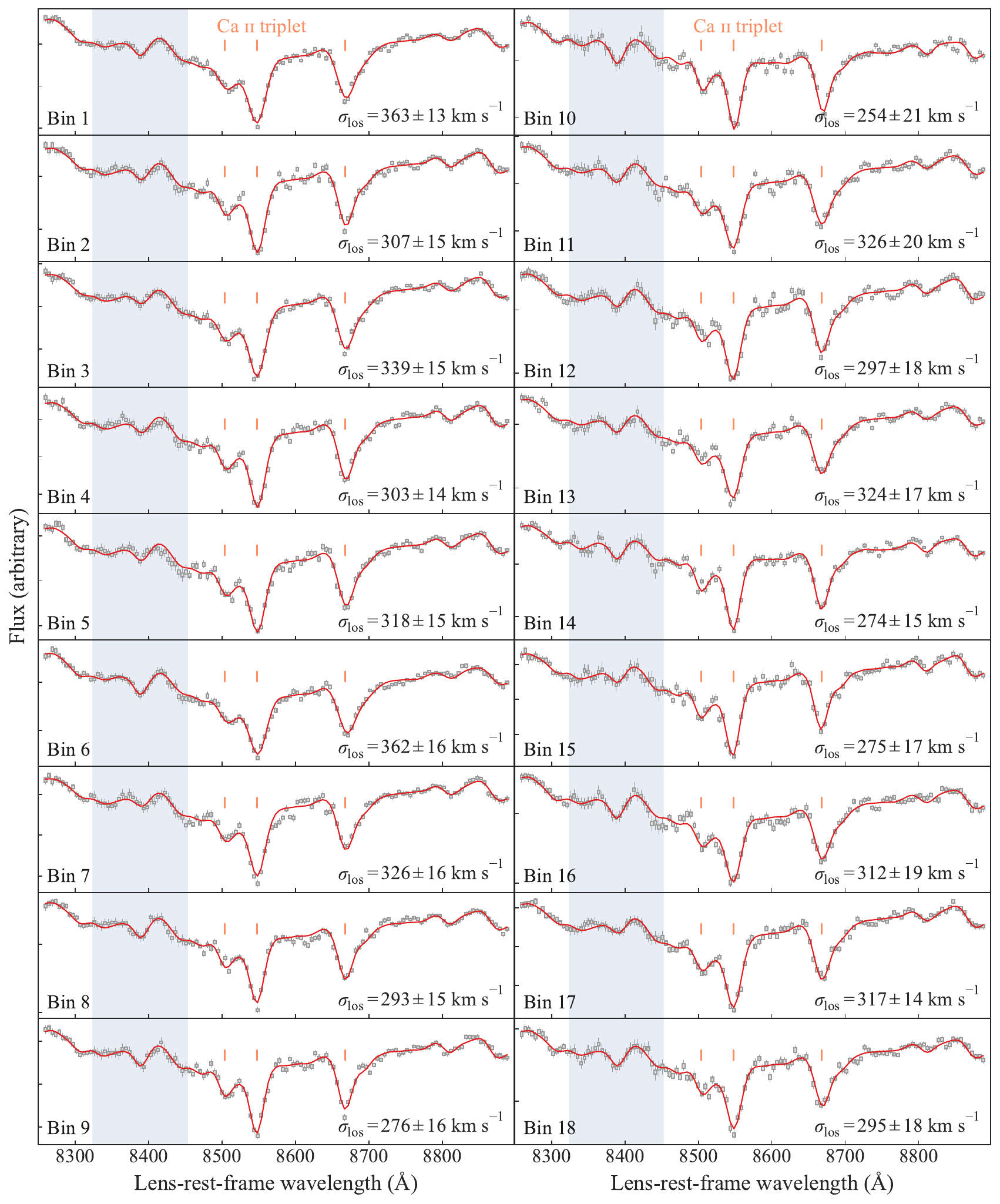}
	\caption{ \label{fig:ppxf_fits_1}
	Individual kinematic fits for the first 18 out of the 32 Voronoi-binned spectra. The best-fit gas lines are subtracted for better visualization of the \cat\ features (marked with orange lines). The blue-shaded region marks where we boosted the noise levels around the H$\alpha$ and N \textsc{ii} lines. The gray bars represent the observed data, with the width marking the bin width and the height marking the original $1\sigma$ noise level. The vertical gray lines represent the boosted 1$\sigma$ noise level. These $1\sigma$ noise levels for each bin are estimated from random sampling given the covariance matrix. The measured velocity dispersion $\sigma_{\rm los}$ is also annotated in each panel.
	}
\end{figure*}

\begin{figure*}
	\includegraphics[width=\textwidth]{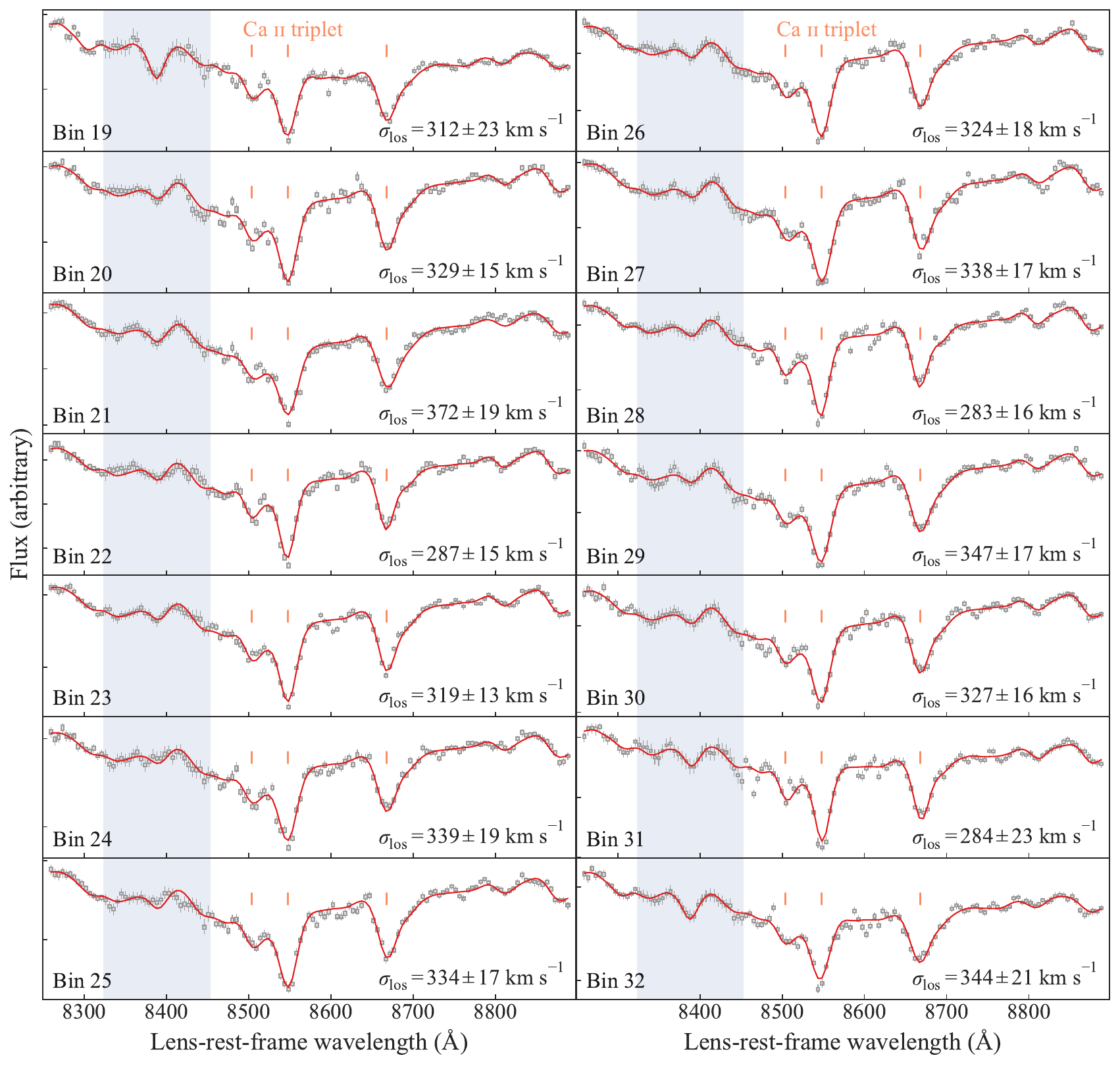}
	\caption{\label{fig:ppxf_fits_2}
		Individual kinematic fits for the last 14 out of the 32 Voronoi-binned spectra. The best-fit gas lines are subtracted for better visualization of the \cat\ features (marked with orange lines). The blue-shaded region marks where we boosted the noise levels around the H$\alpha$ and N \textsc{ii} lines. The gray bars represent the observed data, with the width marking the bin width and the height marking the original $1\sigma$ noise level. The vertical gray lines represent the boosted 1$\sigma$ noise level. These $1\sigma$ noise levels for each bin are estimated from random sampling given the covariance matrix. The measured velocity dispersion $\sigma_{\rm los}$ is also annotated in each panel.
		}
\end{figure*}

In this appendix, we illustrate the kinematic fits to all the individual Voronoi-binned spectra in Figs.~\ref{fig:ppxf_fits_1} and \ref{fig:ppxf_fits_2}.	
	
\end{document}